\documentclass[
reprint,
amsmath, 
amssymb, 
aps, 
superscriptaddress,
pra,
]
{revtex4-2}
\usepackage{graphicx}
\usepackage{dcolumn}
\usepackage{bm}
\usepackage{braket} 
\usepackage{hyperref}
\usepackage{subfigure} 
\usepackage{nccmath}
\usepackage{amsthm} 
\usepackage{mathrsfs} 
\usepackage{xfrac} 
\usepackage{pifont}
\usepackage[export]{adjustbox}
\usepackage{bbm} 
\usepackage[utf8]{inputenc} 
\usepackage[normalem]{ulem}
\usepackage[usenames,dvipsnames,svgnames,table]{xcolor}
\usepackage[sort&compress]{natbib}
\usepackage{color}
\usepackage{longtable}
 
\usepackage{commath}
\usepackage{multirow}
\usepackage{indentfirst}
\usepackage[ruled,vlined]{algorithm2e}
\usepackage{amsmath}
\usepackage{amsfonts}
\usepackage{amssymb}

\theoremstyle{definition}

\usepackage{minitoc}
\usepackage[toc]{appendix}

\begin{document}
\title{Towards a Linear-Ramp QAOA protocol: Evidence of a scaling advantage in solving some combinatorial optimization problems}

\author{J. A. Monta\~nez-Barrera}
 	\altaffiliation{Corresponding author: J. A. Monta\~nez-Barrera; j.montanez-barrera@fz-juelich.de}
	\affiliation{Jülich Supercomputing Centre, Institute for Advanced Simulation, Forschungszentrum Jülich, 52425 Jülich, Germany}
\author{Kristel Michielsen}
	\affiliation{Jülich Supercomputing Centre, Institute for Advanced Simulation, Forschungszentrum Jülich, 52425 Jülich, Germany}
	\affiliation{AIDAS, 52425 Jülich, Germany}
	\affiliation{RWTH Aachen University, 52056 Aachen, Germany}

\begin{abstract}

The Quantum Approximate Optimization Algorithm (QAOA) is a promising algorithm for solving combinatorial optimization problems (COPs), with performance governed by variational parameters \(\{\gamma_i, \beta_i\}_{i=0}^{p-1}\). While most prior work has focused on classically optimizing these parameters, we demonstrate that fixed linear ramp schedules, linear ramp QAOA (LR-QAOA), can efficiently approximate optimal solutions across diverse COPs. Simulations with up to \(N_q=42\) qubits and \(p=400\) layers suggest that the success probability scales as \(P(x^*) \approx 2^{-\eta(p) N_q + C}\), where \(\eta(p)\) decreases with increasing \(p\). For example, in Weighted Maxcut instances, \(\eta(10) = 0.22\) improves to \(\eta(100) = 0.05\). Comparisons with classical algorithms, including simulated annealing, Tabu Search, and branch-and-bound, show a scaling advantage for LR-QAOA. We show results of LR-QAOA on multiple QPUs (IonQ, Quantinuum, IBM) with up to \(N_q = 109\) qubits, \(p=100\), and circuits requiring 21,200 CNOT gates. Finally, we present a noise model based on two-qubit gate counts that accurately reproduces the experimental behavior of LR-QAOA.

\begin{description}
	\vspace{0.2cm}
	\item[Keywords] Combinatorial Optimization, QAOA, LR-QAOA, traveling salesman, maximal independent set, bin packing, portfolio optimization, weighted Maxcut, knapsack, MAX-3-SAT.
\end{description}

\end{abstract}
\maketitle

\section{Introduction}
Finding high-quality solutions for COPs is perceived as one of the main applications of quantum computation in the near future. In the gate-based regime, QAOA \cite{Farhi2014} has become one of the most studied quantum algorithms for solving COPs. There are different factors for the extensive study of QAOA. Firstly, parametric unitary gates can effectively represent the Hamiltonian of the COPs, where the ground state encodes the optimal solution of the problem \cite{Lucas2014, Kochenberger2014}. Moreover, QAOA has a performance guarantee in the limit of infinite layers resembling the quantum adiabatic algorithm \cite{Farhi2000, Farhi2014}. Additionally, QAOA needs fewer resources (e.g., number of gates and qubits) compared to other quantum algorithms and can be tested on current state-of-the-art quantum hardware \cite{Harrigan2021, Niroula2022, Shaydulin2023}. Furthermore, classical methods find it hard to solve large instances of COPs with practical applications \cite{Ohzeki2020}, and therefore, finding alternative ways to solve them is needed. Ultimately, the goal of quantum optimization algorithms, as exemplified by QAOA, is to demonstrate advantages in solving optimization problems, be it in terms of energy efficiency, time-to-solution (TTS), or solution quality compared to classical methods.

In the simplest version of QAOA, the cost Hamiltonian of a combinatorial optimization problem is encoded in a parametric unitary gate along with a ``mixer", a second parametric unitary gate that does not commute with the first unitary gate. In this context, parameters $\gamma = [\gamma_0, ...,\gamma_{p-1}]$ and $\beta = [\beta_0, ...,\beta_{p-1}]$ for the cost and mixer Hamiltonians, respectively, are adjusted to minimize the expectation value of the cost Hamiltonian for $i=0, ..., p-1$ layers of QAOA. Since its conception, a classical algorithm was suggested to find the QAOA $\gamma$ and $\beta$ parameters \cite{Farhi2014}. This makes QAOA fall in the category of variational quantum algorithms (VQA) \cite{Cerezo2021}. However, these algorithms have exhibited a limited/poor performance advantage as the classical optimization part finds it hard to escape local minima when searching for $\gamma$ and $\beta$ parameters \cite{Bittel2021, Kremenetski2021}. The barren plateau is another challenge in QAOA. It refers to regions in the cost function landscape where the gradient is nearly zero, making it hard to find QAOA parameters via gradient-based optimization \cite{Larocca_2022}.

Modest progress has been made by considering QAOA as a VQA,  with major studies conducted in regimes of a few qubits and shallow circuits \cite{Harrigan2021}. Deep QAOA circuits, when viewed through the lens of VQA, lead to a pessimistic conclusion regarding their universal applicability \cite{Kossmann2022}. Moreover, implementations on real hardware face an even greater challenge; the noise inherent in current quantum devices makes the search for the minima of the objective function unfeasible after only a few QAOA layers \cite{Zhou2020, Kossmann2022}.

Alternatively to this methodology, one can fix the $\gamma$ and $\beta$ parameters following some protocol, similar to what quantum annealing (QA) does \cite{Apolloni1989, DeFalco2011}. In this scenario, no further classical optimization is needed. 

Initial evidence supporting the effectiveness of fixed-parameter QAOA was presented by Brandao et al. \cite{Brandao2018}. They demonstrated that fixed parameters exhibit consistent performance regardless of the problem or problem size, suggesting the potential reduction of the outer loop of classical optimization in QAOA. 

Various protocols have been proposed to fix these parameters. In \cite{Kremenetski2021}, Krementski et.~al.~found a set of QAOA parameters with consistent performance to find optimal solutions using a fixed LR-QAOA protocol. They tested this methodology using the Hamiltonian of different molecules, an Ising Hamiltonian, and the 3-SAT problem for intermediate to large $p$. Another attempt to fix the QAOA protocol is proposed in \cite{Willsch2022}, in which the authors presented QAOA as a second-order time discretization of QA referred to as approximate quantum annealing (AQA). In \cite{hess2024effectiveembeddingintegerlinear} is proposed the Trotterized adiabatic evolution (TAE), an idea similar to AQA but using a fixed sinusoidal schedule. LR-QAOA can be considered as an AQA protocol with a linear annealing schedule.

In \cite{Montanez-Barrera2024}, we proposed fixed schedules transferring optimal $\gamma$ and $\beta$ parameters between different COPs. We found that sometimes $\gamma$ and $\beta$ parameters that work well for a COP in the form of Eq. \ref{Eq:1} give good results on other COPs with different structures. Specifically, we found that parameters optimized for the bin packing problem (BPP) can be translated to Maxcut, Maximal independent set (MIS), portfolio optimization (PO), and travelling salesman problem (TSP), giving a quadratic speedup over random guessing on all of them. This suggests that there are effective QAOA protocols that work for different problems. This information led us to the results of the present work. 

Recently, Kremenetski et.~al.~have explained the behavior of LR-QAOA and, in general, of the gradually changing schedules using the discrete adiabatic theorem involving a wrap-around phenomenon \cite{Kremenetski2023}.

In this paper, we extend the study of LR-QAOA schedules to different COPs, presenting numerical and experimental evidence that LR-QAOA constitutes an effective QAOA protocol, i.e., the set of $\gamma$ and $\beta$ parameters from a linear ramp schedule works effectively for many problems and problem sizes in combinatorial optimization. We test this protocol using MIS, BPP, TSP, Maxcut, weighted maximum cut (WMaxcut), 3-regular graph maximum cut (3-Maxcut), Knapsack (KP), PO, maximum 2 Boolean satisfiability problem (Max-2-SAT), and maximum 3-SAT (Max-3-SAT). We use random instances of these COPs with problem sizes ranging from 4 to 42 qubits and $p$ from 3 to 400. For large problems, we simulate them using JUQCS–G software \cite{Willsch2022} on JUWELS Booster, a cluster of 3744 NVIDIA A100 Tensor Core GPUs, integrated into the modular supercomputer JUWELS \cite{Krause2019, JuwelsClusterBooster}. 

In these cases, the average probability of success over the 100 random instances seems to follow a scaling that can be described by $probability(x^*) = 2^{-\eta(p) N_q + C}$ for a $\eta(p)$ decreasing with $p$ and a constant $C$. We extend the analysis to fully connected random WMaxcut. The WMaxcut is both NP-Hard \cite{Gutin2021} and APX-Hard \cite{Paradimitriou1991} problems.

We find a scaling improvement in terms of the time-to-solution (TTS) \cite{Zhou2020} when using LR-QAOA compared to SA, TABU search, and B\&B for solving random instances of WMaxcut. This evidence complements the recent findings in \cite{Shaydulin2023b} where a scaling advantage is observed in a fixed QAOA protocol for solving a classical intractable COP known as low autocorrelation binary sequences (LABS) and on k-SAT problems \cite{Boulebnane2024}.

We extend the analysis to real quantum hardware. Using IonQ Aria ({\it ionq\_aria}), Quantinuum H2-1 ({\it quantinuum\_H2}) \cite{Moses2023}, IBM Brisbane ({\it ibm\_brisbane}), IBM Kyoto ({\it ibm\_kyoto}), IBM Osaka ({\it ibm\_osaka}), and IBM fez ({\it ibm\_fez}), we run WMaxcut problems ranging from 5 to 109 qubits and $p$ from 3 to 100. We find that there is an effective number of layers, $p_{\mathrm{eff}}$, for which the best performance is obtained using each device. 

In the case of IBM devices $p_{\mathrm{eff}} = 10$, on {\it ionq\_aria} $p_{\mathrm{eff}}=10$, and on {\it quantinuum\_H2} $p_{\mathrm{eff}} = 50$. Remarkably, for the largest problem size, 109 qubits and $p=100$, we observe that LR-QAOA still possesses an improvement over random sampling in {\it ibm\_kyoto} and {\it ibm\_osaka}. For a comparative analysis between the different vendors, we test a 25-qubit WMaxcut problem on them, {\it quantinuum\_H2} gives the best performance with a probability($x^*$) $=0.08$ at $p=50$.

We present a noise model of LR-QAOA that fits depolarizing noise simulation and experiments on ibm\_fez and an emulator of Quantinuum H1-1. The noise model depends only on the number of 2-qubit gates and a noise parameter associated with the QPU. We observe that there is an interplay between the noise pushing the system towards a maximally mixed state and LR-QAOA driving the system towards the minimum energy of the cost Hamiltonian.

\section{Methods}\label{Sec:Methods}
In this section, we describe the LR-QAOA, some properties of LR-QAOA, the combinatorial optimization problems used, the classical solvers used to compare scaling properties, and experimental details on real quantum hardware.
\subsection{LR-QAOA}

\begin{figure*}[!tbh]
\centering
\includegraphics[width=17.5cm]{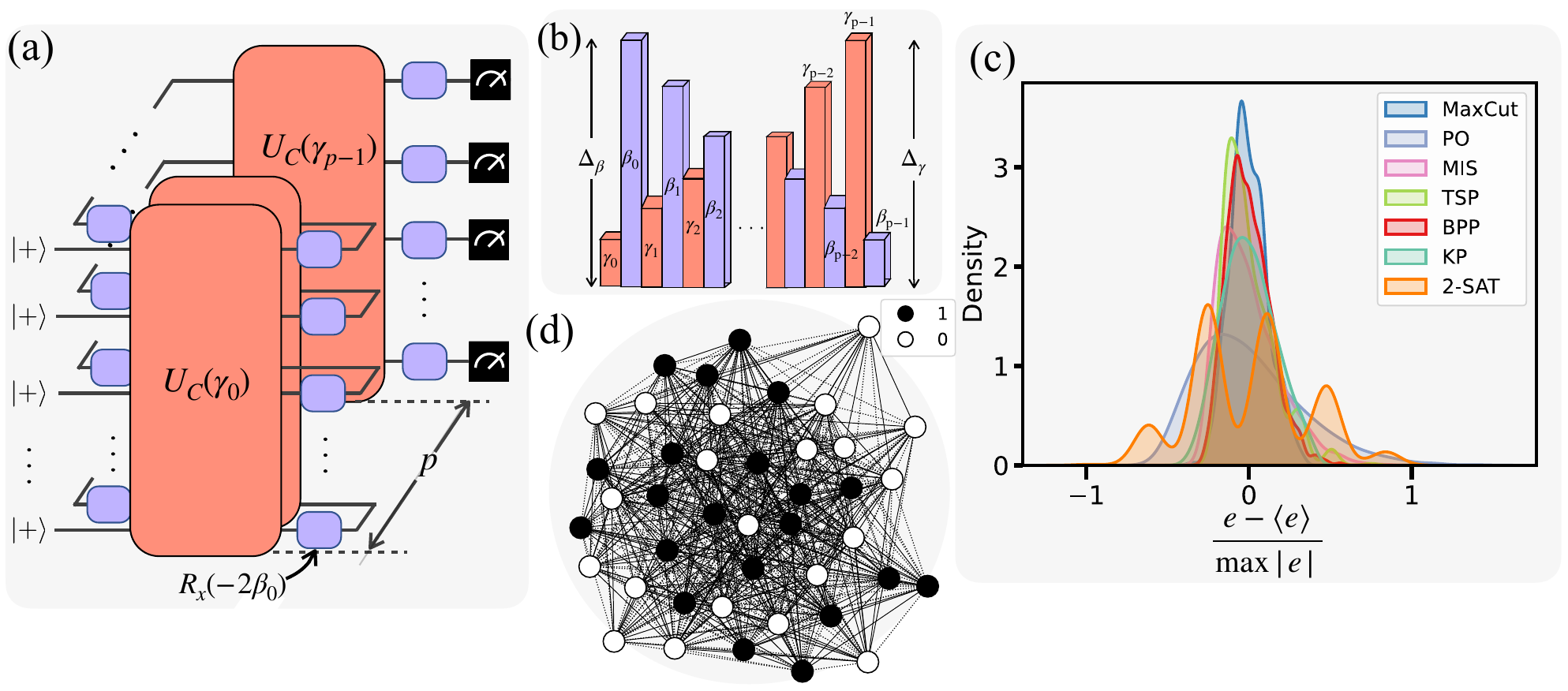}
\caption{\label{Fig:linear_ramp_schedule}(a) Quantum circuit of the QAOA algorithm, (b) LR-QAOA schedule, (c) Density vs the normalized eigenvalue distribution for the different COPs with $e$ representing the eigenvalue. All the distributions are for 10-qubit problems except BPP and TSP, with 12-qubit and 9-qubit problems, respectively.  (d) An optimal solution for one of the 42-qubit WMaxcut problems using $p=50$ LR-QAOA. Dashed lines represent cuts, black (white) vertices qubits in 1 (0) state. At the end of the algorithm, the probability of finding the maximum cut is 32\%.}
\end{figure*}

QAOA consists of alternating layers that encode the problem of interest along with a mixer element in charge of amplifying solutions with low energy. In this case, the COP cost Hamiltonian is given by

\begin{equation}\label{Eq:1}
H_C = \sum_i h_i \sigma_z^i + \sum_{i, j > i} J_{ij} \sigma_z^i \sigma_z^j, 
\end{equation}
where $\sigma_z^i$ is the Pauli-z term of qubit i, and $h_i$ and $J_{ij}$ are coefficients associated with the problem. Usually, $H_C$ is derived from the quadratic unconstrained binary optimization (QUBO) formulation \cite{Lucas2014, Montanez-Barrera2024, Montanez-Barrera2022}. The QUBO to $H_C$ transformation usually includes a constant term that does not affect the QAOA formulation and is left out for simplicity. $H_C$ is encoded into a parametric unitary gate given by

\begin{equation}\label{UC}
    U_C(H_C, \gamma_i)=e^{-j \gamma_i H_C},
\end{equation}
 where $\gamma_i$ is a parameter that in our case comes from the linear ramp schedule. Following this, in every second part of a layer, a unitary operator is applied, given by 

\begin{equation}\label{UB}
    U(H_B, \beta_i)=e^{j \beta_i H_B},
\end{equation}
where $\beta_i$ is taken from the linear ramp schedule and $H_B = \sum_{i=0}^{N_q-1} \sigma_i^x$ with $\sigma_i^x$ the Pauli-x term of qubit $i$. The general QAOA circuit is shown in Fig.~\ref{Fig:linear_ramp_schedule}-(a). Here, $R_X(-2\beta_i) = e^{j\beta_i \sigma^x}$, $p$ is the number of repetitions of the unitary gates of Eqs.~\ref{UC} and \ref{UB}, and the initial state is a superposition state $| + \rangle^{\otimes N_q}$. Repeated preparation and measurement of the final QAOA state yields a set of candidate solution samples, which are expected to give the optimal solution or some low-energy solution.

In Fig ~\ref{Fig:linear_ramp_schedule}-(b), we show the LR-QAOA protocol. It is characterized by three parameters $\Delta_\beta$, $\Delta_\gamma$, and the number of layers $p$. The $\beta_i$ and $\gamma_i$ parameters are given by 

\begin{equation}
\beta_i = \left(1-\frac{i}{p}\right)\Delta_\beta\ \ \mathrm{and} \ \
\gamma_i = \frac{i+1}{p}\Delta_\gamma,
\end{equation}
for $i=0, ..., p-1$. For our simulations, we scan over a set of $\Delta_{\gamma}$ and $\Delta_{\beta}$ from one problem at each problem size and use the best value over the remaining cases. For the experimental results, we use $\Delta_\beta = 0.3$ and $\Delta_\gamma = 0.6$.

\subsection{Properties of LR-QAOA}\label{Sec:Properties-LR-QAOA}

\begin{figure*}[!tbh]
\centering
\includegraphics[width=17.5cm]{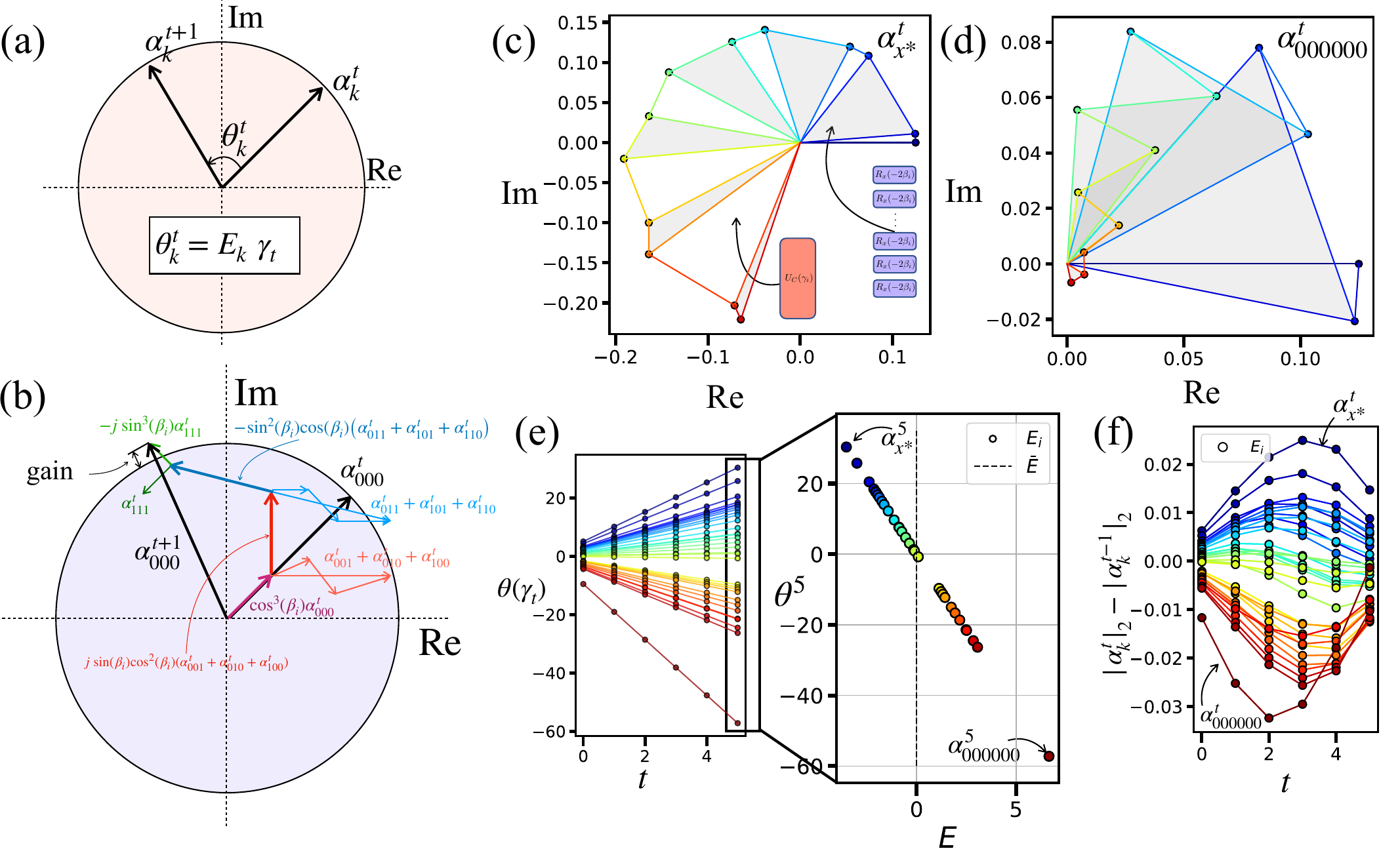}
\caption{\label{Fig:sv_evolution}LR-QAOA from the point of view of individual time steps. (a) The action of the $U_C(\gamma_t)$ gate on the state $k$ at time step $t$, (b) evolution of the $|000\rangle$ amplitude after the application of $U_B(\beta_t)$ for a 3-qubit system, (c) evolution of the optimal solution, $x^*$, in a 6-qubit WMaxcut problem. The gray (white) triangles are a time-step evolution due to $U_B(\gamma_t)$ ($U_C(\gamma_t)$). Line colors represent the time steps being blue (red) $t=0$ ($t=5$) step. (d) Evolution of the worst solution for the 6-qubit WMaxcut problem. (e)-left LR-QAOA $\gamma$ rotations at each layer for each state. Positive angles refer to counterclockwise rotations. Colors represent the energy of the state, with darker blue (red) closer to the optimal (worst) solution of the problem. (e)-Right last layer rotation in LR-QAOA vs. the energy, following Eq~\ref{angle_rotation}. (f) Amplitude gain evolution of the states after each $U_B(\beta_t)$ for the 6-qubit WMaxcut problem.}
\end{figure*}

The QAOA evolution is usually presented from the point of view of the expectation value of the cost Hamiltonian \cite{Farhi2014, Hadfield2023, Kremenetski2023}. In this section, we present a framework where the evolution under LR-QAOA is seen from the point of view of the individual amplitudes of all possible states in a COP. The state vector that describes the evolution of $probability(x^*)$ of a COP is given by

\begin{equation}
|\psi_t\rangle = \sum_{k=0}^{2^{N_q} -1} \alpha^t_{k} |k\rangle,
\end{equation} 

where $t$ is some step in the QAOA algorithm, $k$ is the state in the computational basis, and $\alpha_k^t$ the amplitude of $|k\rangle$ at time $t$.

The unitary transformation induced by $U_C(\gamma_t)$, $|\psi_{t+1}\rangle = U_C(\gamma_t)|\psi_{t}\rangle$, produces a rotation in the complex plane for every state given by

\begin{equation}
\alpha_{k}^{t+1} = e^{j\theta_{k}^t} \alpha_{k}^{t},
\end{equation}

\begin{equation}\label{angle_rotation}
    \theta_{k}^t = E_k \gamma_t ,
\end{equation}

where $E_k = \langle k| H_c |k\rangle$. This evolution is shown in Fig. \ref{Fig:sv_evolution} (a). 
Eq.~\ref{angle_rotation} explains why the amplitude amplification is proportional to the energy of a given solution. Negative energies are rotated counterclockwise with the rotation proportional to their energies. This can be seen in Fig. \ref{Fig:sv_evolution}-(e).

The change by $U_B(\beta_t)$, $|\psi_{t+1}\rangle = U_B(\beta_t)|\psi_{t}\rangle$, is more complex and depends on the Hamming distance of the given state to the other states. This operator is responsible for the change in energy and produces an interference pattern that exploits the $U_C(\gamma_t)$ effect. It is described by

\begin{equation}\label{Eq:alpha}
    \alpha^{t+1}_{k} = \sum_{l=0}^{2^{N_q - 1}} \left(\cos(\beta_t)\right)^{N_q - k \cdot l}\left(j \sin(\beta_t)\right)^{k\cdot l} \alpha^{t}_l,
\end{equation}
with

\begin{equation}\label{Eq:Hamm}
    k\cdot l = \sum_{m=0}^{N_q-1} (k_m\oplus l_m),
\end{equation}
where $k$ and $l$ are states in the computational basis. Equation~\ref{Eq:Hamm} gives the Hamming distance between the states $k$ and $l$. See Supplementary Note 5 for a detailed derivation of Eq.~\ref{Eq:alpha}. For example, in a 3-qubit system, the evolution of $\alpha^t_{000}$ is given by

\begin{align*}
\alpha_{000}^{t+1}  = & \ \langle 000 | U_B(\beta_t) |\psi_t\rangle\\
=& \ \cos^3(\beta_t)\alpha_{000}^t \\ & +  j\sin(\beta_t)\cos^2(\beta_t)\left( \alpha_{001}^t + \alpha_{010}^t + \alpha_{100}^t\right) \\ & - \sin^2(\beta_t)\cos(\beta_t)\left( \alpha_{011}^t + \alpha_{101}^t + \alpha_{110}^t\right)\\ & -j \sin^3(\beta_t)\alpha_{111}^t.
\end{align*}
A schematic representation of how the $U_B(\beta_t)$ induces an evolution of $\alpha_{000}$ is shown in Fig. \ref{Fig:sv_evolution}-(b). Here, $U_B(\beta_t)$ changes the amplitude and direction of $\alpha_{000}$ using the information of $\alpha_{000}$ and the other states. The Hamming distance indicates how the amplitudes are grouped. For example, the effective vector $r_1 = (\alpha_{001} + \alpha_{010}+\alpha_{100})$ of states with Hamming distance 1, contribute to $\alpha_{000}$ after a $\pi/2$ rotation and a rescaling given by $\sin(\beta_t) \cos^2(\beta_t)$.

In Fig. \ref{Fig:sv_evolution}-(c) is shown the evolution of the optimal solution, $x^*$, of a 6-qubit WMaxcut problem for the $U_C(\gamma_t)$ and $U_B(\beta_{t})$ steps for $t \in \{ 0,..., Nq-1\}$. In Fig.\ref{Fig:sv_evolution}-(d) shows the same evolution but for the state with the lowest energy. In this case, the evolution of the eigenvalue due to $U_C(\gamma_t)$ goes in the opposite direction to the evolution of $U_B(\beta_t)$ producing the desired effect of interference. Fig.\ref{Fig:sv_evolution}-(e) shows the angle of rotation of the white triangles, i.e., the rotation due to $\theta(\gamma_t)$. The gain in the amplitude of the $\alpha_k^t$ at each time step after the unitary evolution $U_B(\beta_t)$ is shown in Fig. \ref{Fig:sv_evolution}-(f).

\subsection{COPs}\label{Sec:COPS}

A detailed description of some COPs used in this work can be found in the appendix of \cite{Montanez-Barrera2024}, and for the Max-3-SAT is presented in the Supplementary Note 6. For them, we use a normalization technique described in Sec.~\ref{SubSec:H}. We pick 5 random instances for different problem sizes. For the TSP, we use instances with 3, 4, 5, and 6 cities (9, 16, 25, and 36 qubits), where the distances between cities are symmetric and randomly chosen from a uniform distribution with values between 0.1 to 1.1. In the BPP, we consider scenarios involving 3, 4, 5, and 6 items (12, 20, 30, and 42 qubits). The weight of each item is randomly chosen from 1 to 10, and 20 is the maximum weight of the bins. The WMaxcut, 3-Maxcut, MIS, and PO problem sizes are given by the number of qubits and chosen to be 20, 25, 30, 35, and 40.

For WMaxcut problem simulations, we use randomly weighted edges with weights chosen uniformly between 0 and 1 and edge density, $E_d = 0.7$. One of these cases with its optimal solution is presented in Fig. \ref{Fig:linear_ramp_schedule}-(d). To test the scaling of LR-QAOA, we use a fully connected random WMaxcut with weights chosen from a uniform discrete distribution from 0 to 1000 in steps of 1. For MIS, edges between nodes are randomly selected with $E_d = 0.4$. For KP problems, item values range from 5 to 63, weights from 1 to 20, and the maximum weight is set to half of the sum of all weights. Finally, for PO, correlation matrix values are chosen from $[-0.1, 0, 0.1, 0.2]$, asset costs varying between 0.5 and 1.5, and the budget is set to half of the total asset cost.

For the inequality constraints in the KP, PO, and BPP, we use the unbalanced penalization approach \cite{Montanez-Barrera2022, Montanez-Barrera2023}.  In this approach, two penalty terms in the QUBO are tuned following the characteristics of the inequality constraints and the objective function. Consequently, any variation in the parameter range necessitates a re-tuning of the penalty terms to maintain performance. For the $probability(x^*)$ using unbalanced penalization, our focus is on finding the ground state of the cost Hamiltonian, since we are interested in knowing the LR-QAOA performance in finding the ground state of the Hamiltonian and there is no guarantee that the optimal solution of the original problem is encoded in the ground state of the Hamiltonian (see also the discussion in \cite{Montanez-Barrera2022}). 

From the problems tested, MIS, BPP, TSP, Maxcut, WMaxcut, KP, PO, and Max-3-SAT are NP-hard \cite{Lucas2014, Paradimitriou1991}, with varying structural properties and practical solution approaches. Some of them admit effective approximation schemes and are commonly addressed using heuristics or dynamic programming in restricted cases. In particular, MIS and PO have been included in a list of 10 classical hard problems that might benefit from quantum algorithms \cite{koch2025quantumoptimizationbenchmarklibrary}.

We use the $probability(x^*)$ as a metric of the performance for the different COPs. Here, $x^*$ represents the set of optimal bitstrings of the problem's Hamiltonian. Additionally, we use the approximation ratio for the Maxcut and its variations. The approximation ratio is given by 

\begin{equation}
r = \frac{\sum_{i=1}^{n} C(x_i)/n}{C(x^*)},
\label{eq:r}
\end{equation}

\begin{equation}
C(x) = \sum_{k,l>k}^{N_q} w_{kl}(x_k + x_l - 2x_kx_l),
\end{equation}
where $n$ is the number of samples, $x_i$ the $ith$ bitstring obtained from LR-QAOA, and  $C(x)$ is the cost function of WMaxcut, $x^*$ is the optimal bitstring, $C(x^*)$ is the maximum cut, $w_{kl}$ is the weight of the edge between nodes $k$ and $l$, and $x_k$ is the kth position of the $x$ bitstring.

Figure \ref{Fig:linear_ramp_schedule}-(c) presents examples of the eigenvalue distribution of the Hamiltonian for different COPs. In the scenario of large-scale problems, the distribution of eigenvalues tends to converge to a normal distribution \cite{Wald1944}.

\subsection{Hamiltonian normalization}\label{SubSec:H}

The Hamiltonian normalization is one important step in LR-QAOA. As we show, every eigenvalue rotates accordingly to Eq. \ref{angle_rotation}, which means that the normalization limits the rotation angle, fixing the {\it ridge region} to a specific location in the performance diagram \cite{Kremenetski2021} (See Supplementary Note 7). The general form of the COP's Ising Hamiltonian is given by

\begin{equation}\label{IsingH}
H_c(\mathrm{z}) = \frac{1}{\max\{|J_{ij}|\}}\left(\sum_{i=0}^{n-1}\sum_{j>i}^{n-1} J_{ij} z_i z_j + \sum_{i=0}^{n-1} h_{i} z_i + \text{O}\right),
\end{equation} 
where $J_{ij}$ and $h_i$ are real coefficients that represent the COP, and the offset, O, is a constant value. Since O does not affect the location of the optimal solution, it can be left out for the sake of simplicity. There are different ways of normalizing the Hamiltonian, we identify two, normalizing by $\max\{|J_{ij}|\}$ or $\max\{|h_{i}, J_{ij}|\}$, and use them on each problem. We select the one with the best results in terms of $probability(x^*)$. We find that the $\max\{|J_{ij}|\}$ strategy improves faster the $probability(x^*)$ while $\max\{|h_{i}, J_{ij}|\}$ improves optimal and suboptimal energies. For the results presented, we choose to normalize the Hamiltonian by $\max\{|J_{ij}|\} \ \forall \ i>j \in {0,.., n-1}$ for almost all the cases except MIS where we use $\max\{|h_{i}|\} \ \forall \ i \in {0,.., n-1}$. 
\subsection{Classical solvers}

To assess the performance of LR-QAOA, we compare its scalability to simulated annealing (SA) \cite{Kirkpatrick1983}, TABU search \cite{Palubeckis2004}, and CPLEX's spatial B\&B \cite{Bliek2014SolvingMQ}. We selected TABU search because it has been shown to outperform other solvers in finding optimal solutions to Maxcut problems \cite{Dunning2018}. The improved performance in the TABU search can be attributed to a TABU list that prevents revisiting previous solutions and therefore mitigates local minimum problems. We use time-to-solution (TTS) as a metric to compare the resources needed to find the optimal solutions to fully connected WMaxcut. The $TTS$ is given by 

\begin{equation}\label{Eq:TTS}
    TTS_{p_d} = T\frac{\ln(1-p_d)}{\ln(1-probability(x^*))},
\end{equation}
where T is the time needed to get one sample, $p_d=0.99$ is the target probability, i.e., the confidence level that the optimal solution is sampled at least once with 99\% certainty. T in SA and TABU depend on the number of sweeps, with 1 sweep representing a full update cycle over all variables. In experiments, we vary the number of sweeps from 50 to 500.

For SA, we use the \texttt{dwave-neal} \cite{neal_sampler} and for TABU we use \texttt{dwave-tabu} \cite{dwave_tabu}, both performant C++-based libraries that use Python as an interface. In the case of the CPLEX solver, we use \texttt{docplex} \cite{docplex} Python interface of CPLEX. All the algorithms run on a MacBook Pro equipped with an Apple M1 chip. The code used to run the given cases can be found at \cite{LR-QAOA}.

The case of LR-QAOA, the $T=t_{2q}(2N_q+2)p$ with $t_{2q}$ is the 2-qubit gate time, and the time to execute one layer of QAOA scales as $O(2N_q+2)$ based on a flexible scheme that can be run in a 1D chain of qubits \cite{klaver2024}. The $t_g$ for most superconducting-based QPUs is on the order of nanoseconds.

\subsection{Noise model} \label{Sec:noise}

At the instruction level, the main source of noise in digital quantum computers comes from the 2-qubit entangling gates \cite{Pascuzzi2022}. Thus, we use a depolarizing noise channel in the 2-qubit gates of the LR-QAOA protocol. This channel is given by

\begin{equation}\label{Eq:depolarizing}
\mathcal{E}[\rho] = (1 - \lambda) \rho + \lambda \frac{I}{4},
\end{equation}
where $\lambda$ is the depolarizing error parameter, $I$ is a $4 \times 4$ identity matrix, and $\rho$ is the density matrix of the 2-qubit system. In general, the action of a 2-qubit gate on a general density matrix can be expressed by 

\begin{equation}
\mathcal{E}_{ij}[\rho] = (1- \lambda)U_{2Q}^{ij}\rho U_{2Q}^{ij} + \frac{\lambda}{4} \mathrm{Tr}_{ij}(\rho) \otimes I,    
\end{equation}
where $\mathcal{E}_{ij}$ is the channel acting on $\rho$, $\mathrm{Tr}_{ij}$ is the partial trace over qubits $i$ and $j$, and $U_{2Q}^{ij}$ is the 2-qubit unitary gate. For simplicity, we assume $\lambda$ is the same for all the 2-qubit gates.

To test how noise affects the LR-QAOA solution for a given problem, we use the following relation,

\begin{equation}
    p_{ovl} =\frac{probability(x^*)^{QPU} - probability(x^*)^{r}}{probability(x^*) - probability(x^*)^{r}}, 
\end{equation}

where $p_{ovl}$ is the overlap between the ideal success probability $probability(x^*)$ and the one obtained in the real device, $probability(x^*)^{QPU}$, normalized by the random sampler success probability, $probability(x^*)^{r}$. Additionally, we define the accumulated error in the circuit using

\begin{equation}
    \varepsilon_{acc} = N_{g} \lambda
\end{equation}

where $N_{g}$ is the number of 2-qubit gates involved in the circuit. Using this relation, we find that a model that describes $p_{ovl}$ is 

\begin{equation}\label{Eq:noise}
    p_{ovl} = 2^{-k_0 \varepsilon_{acc}},
\end{equation}

where $k_0$ is a fitting parameter that depends on the problem. In results, we show that this approach can be applied to superconductive and trapped ion-based QPUs, obtaining a good match of experimental results for both devices.

\subsection{Mitigation: Hamming distance 1}\label{A:mitig}

In \cite{Montanez-Barrera2024}, we introduce the mitigation technique used here. This involves applying a bitflip to each position within the output bitstring of samples from a quantum computer, to mitigate single-qubit bitflips errors. The computational overhead of this postprocessing method is $\mathrm{O}(N N_q)$, where N represents the number of samples and $N_q$ is the number of qubits. While this mitigation technique can correct errors coming from the readout of the quantum device, it is also an optimization step that can completely obscure the optimization coming from the LR-QAOA algorithm. Therefore, it is important to compare the results against those obtained from a random sampler using the same mitigation technique, which is included in our main results. The details of our proposed approach are described in Algorithm \ref{Alg:Mitig}.

\begin{algorithm}[h]\label{Alg:Mitig}
    \caption{Sampler mitigation}
    \KwData{bitstring samples S = [$s_0$, ..., $s_{N-1}$]}
    \KwResult{Samples corrected $S_{mitig}$}
    
    Initialization\;
    \For{i=0; i++; $i < N$}{ 
        $E_{best} = \mathrm{Energy}(S[i])$\;
        $s_{best}$ = $S[i]$\;
        \For{j=0; j++; $j < N_q$}{
            $s_{new}$ = $S[i]$\;
            \If{$s_{new}$[j] == 1}{
                $s_{new}$[j] = 0}\
            \Else{$s_{new}$[j] = 1}
            $E_{new}$ = Energy($s_{new}$)\;
            \If{ $E_{new} < E_{best}$}{
                $E_{best}$ = $E_{new}$\;
                $s_{best}$ = $s_{new}$\;
            }
        $S_{mitig}\leftarrow s_{best}$
        }
        
    }
    \Return{$S_{mitig}$}\;
\end{algorithm}

\subsection{Experimental details}\label{exp_details}
We use random fully connected WMaxcut from 5 to 15 qubits. We run experiments on ibm\_fez and H1-1E. For the case of ibm\_fez, we use the parity twine chains (PTC) \cite{dreier2025connectivityaware, klaver2024} strategy to encode the LR-QAOA quantum circuit into a 1D-chain of qubits of the QPU. H1-1E is a 20-qubit emulator of the Quantinuum H1-1 QPU. In these experiments, $\Delta_{\gamma} = \Delta_{\beta} = 0.6$, the number of samples is 1000.

We implement WMaxcut problems using LR-QAOA with $\Delta_\gamma = 0.6$ and $\Delta_\beta=0.3$ on three quantum computing technologies: IonQ Aria a fully connected 25-qubit device based on trapped ions with 2-qubit gate error of 0.4\% and 2-qubit gate speed of $t_{2q} = 600 \mu s$ \cite{ionqaria}, labeled {\it ionq\_aria}, Quantinuum H2-1 (a fully connected 32-qubit device based on trapped ions with a 2-qubit error rate of 0.2$\%$ \cite{Moses2023}, labeled {\it quantinuum\_H2}), and three IBM Eagle superconducting processors \cite{ibmeagle}, 127 transmon qubits with heavy-hex connectivity and 2-qubit median gate error between $0.74$ and $0.95\%$, error per layered gate (EPLG) \cite{McKay2023} between 1.9\% and 3.6\%, and 2-qubit gate speed of $t_{2q} = 0.66 \mu s$, labeled {\it ibm\_brisbane}, {\it ibm\_kyoto} and {\it ibm\_osaka}). 

We perform different experiments to assess the practical performance of quantum technology to solve COPs using LR-QAOA. First, an experiment on {\it ionq\_aria} for a 10-qubit WMaxcut with 70\% of random connections as described in Section \ref{Sec:COPS}, this helps for the sake of comparison with a depolarizing noise model. Additionally, different problems from 5 to 109 qubits were tested on {\it ibm\_brisbane} using a WMaxcut problem with a 1D-chain topology shown in Fig. ~\ref{Fig:r_comparison}-(a). We opt for a simple graph due to constraints posed by noise. Additionally, we provide an experimental comparison across three distinct IBM devices for a 109-qubit WMaxcut problem. Finally, a comparison between {\it ionq\_aria}, {\it ibm\_brisbane}, and {\it quantinuum\_H2} is shown for a 25-qubit WMaxcut problem, Fig.~\ref{Fig:r_comparison}-(b).

The time of execution $t_e$ for the 1D chain topology LR-QAOA protocol can be approximated to that of the 2-qubit gates. This is because single-qubit operations are a minority and their execution time is generally faster than 2-qubit gates. In {\it ionq\_aria} the 2-qubit gates are executed sequentially so $t_e = t_{2q} N_{2q} p$ where $N_{2q}$ is the number of 2-qubit terms in the cost Hamiltonian. For the case of {\it ibm\_brisbane}, the time of execution is $t_e = 2 t_{2q} p$. {\it quantinuum\_H2} can execute 4 2-qubit gates in parallel, hence, the execution time is $t_e =  t_{2q} (N_{2q}/4) p$. The time per 2-qubit gate is $600 \ \mu s$ on {\it ionq\_aria}, and $660 \ ns$ on {\it ibm\_brisbane}. We could not find information about $t_{2q}$ for {\it quantinuum\_H2}, but we assume it is similar to {\it ionq\_aria}. Therefore, a 25-qubit WMaxcut with 1D topology requires $14.4 ms$, $3.6 \ ms$, and $1.32 \ \mu s$ for each layer using {\it ionq\_aria}, {\it quantinuum\_H2}, and {\it ibm\_brisbane}, respectively. For each experiment on IBM devices, {\it ionq\_aria}, and {\it quantinuum\_H2}, we use 10000, 1000, and 50 samples, respectively.

\section{Results}\label{Sec:Results}
\subsection{Simulations}

\begin{figure*}
\centering
\includegraphics[width=17cm]{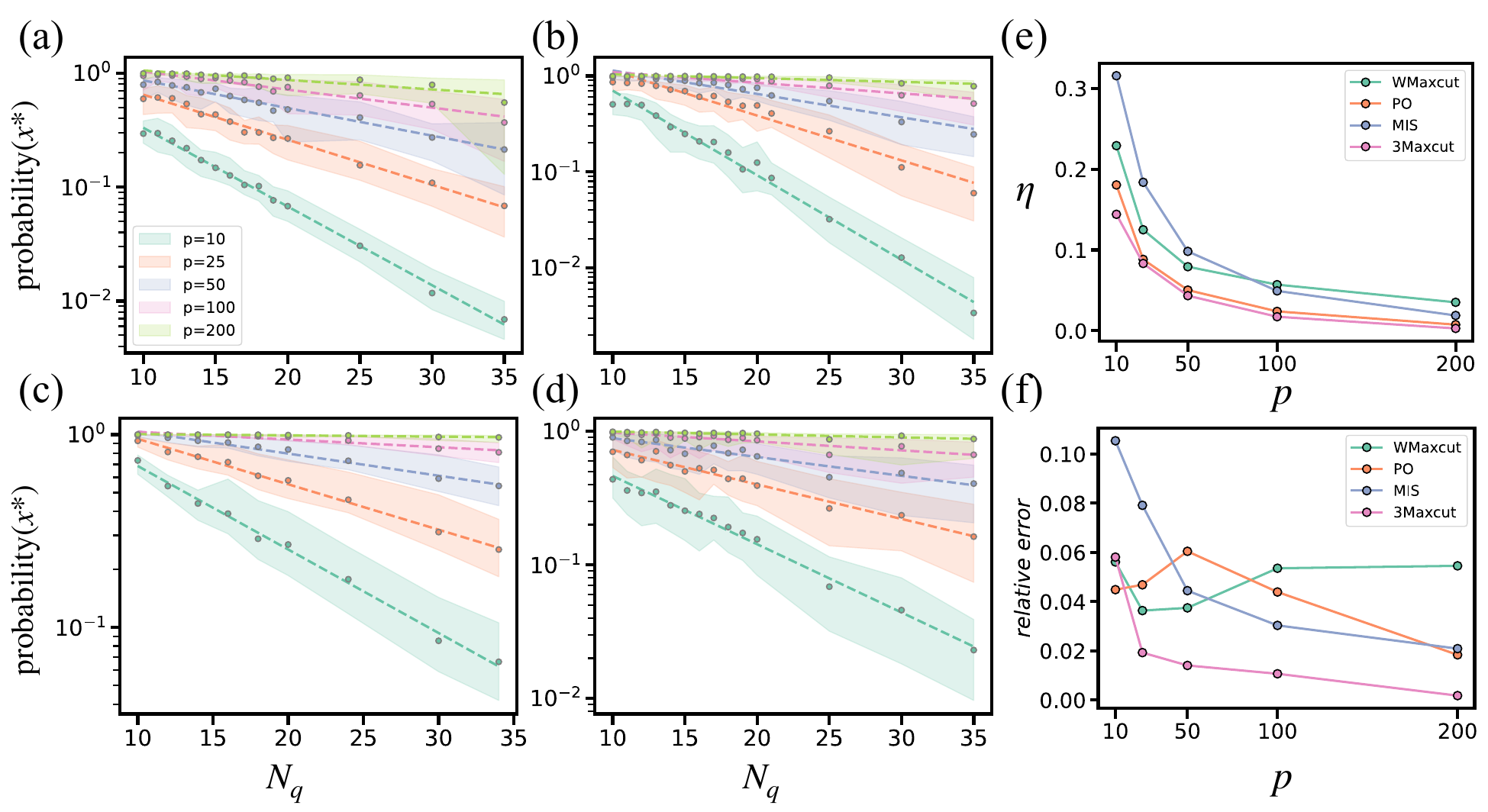}
\caption{\label{Fig:probabilities} Probability of success for 100 random instances of (a) WMaxcut, (b) MIS, (c) 3-Maxcut, and (d) PO. The shaded region represents the quartiles 1 and 3 over the 100 cases. The different colors represent the number of LR-QAOA layers (see legend). Dashed lines represent the conjectured scaling $2^{-\eta(p) N_q + C}$ for each $p$. (e) Fitted $\eta$ vs. number of LR-QAOA layers for the mean value of the problems. (f) Relative error of $probability(x^*)$ calculated using the fitting parameters vs. the number of LR-QAOA layers.}
\end{figure*}

Figures~\ref{Fig:probabilities}-(a)-(d) show the average probability of success for different COPs vs.~the number of qubits. We test 100 random problems with 10 to 35 qubits using the LR-QAOA with $p=10$ to $200$. The $\Delta_\gamma$ and $\Delta_\beta$ values are scanned for each problem size for one instance, and that value is used for the remaining 99 instances. Although this methodology is not optimal, since ideally, individual parameters should be tuned for each case, we adopt this approach to manage the growing simulation costs associated with a large number of qubits. Moreover, this strategy highlights an important feature of LR-QAOA, even without fine-tuning parameters for each instance, the algorithm consistently achieves good performance. In Supplementary Note 1, we present in detail the methodology and the values used. In these problems, we observe that the number of layers affects the scale of the probability of success with a relation that can be described by

\begin{equation}\label{Eq:eta}
probability(x^*) = 2^{-\eta(p) N_q + C},
\end{equation}

with an $\eta(p)$ that is a function of $p$, and $C$ a constant. The perceived scaling still needs to be corroborated at a larger problem size to confirm that the probability of success indeed decreases exponentially. A similar scaling QAOA behaviour has been observed for k-SAT problems in \cite{Boulebnane2024}. If this holds, it means that there is an exponential improvement in LR-QAOA achieved by increasing the number of layers linearly. This does not necessarily mean that the problems are hard and that the best classical solvers for them exhibit the perceived exponential scaling of LR-QAOA. 

Figure~\ref{Fig:probabilities}-(e) shows the fitting values of $\eta(p)$ for each $p$ using the information of Figs~\ref{Fig:probabilities}-(a)-(d). The four models exhibit similar $\eta(p)$ behavior, decaying quickly with the number of layers. The fitted $\eta(p)$ implies an exponential improvement as the number of layers increases. For example, if the WMaxcut scale holds at $N_q = 100$, using LR-QAOA with $p=10$ the $probability(x^*)=2\times10^{-7}$ while $probability(x^*)=0.2$ with $p=200$. Figure~\ref{Fig:probabilities}-(f) shows the relative error of the difference between the value predicted by the fitting curve and the actual values for each $p$, i.e., $\varepsilon = |1 - \frac{2^{-\eta(p) N_q + C}}{probability(x^*)}|$. As the number of layers increases, the error decreases in 3 out of 4 cases and remains below 6\%.

In Supplementary Note 2, we present the details about the scaling factor $\eta(p)$ and its relation with the minimum number of qubits used in the cutoff of the fitting function. The previous version of this paper included a conjecture that the probability of success scales as $probability(x^*) = 2^{-\eta N_q/p}$, with the further evidence of this section, the model of the scale of $probability(x^*)$ has changed. Previous results and extended simulations are presented in Supplementary Note 3. In the next section, we present a comparative analysis of LR-QAOA and several classical solvers.

\subsection{Scaling comparison}\label{Sec:scaling}
\begin{figure*}
\centering
\includegraphics[width=18cm]{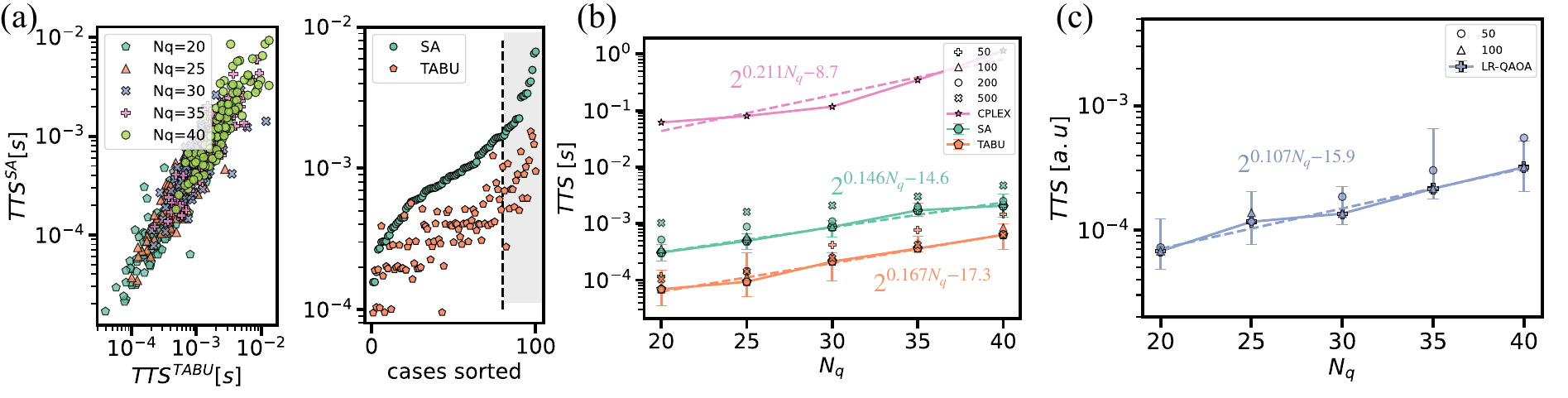}
\caption{\label{Fig:scaling} Scaling comparison in terms of TTS vs. number of qubits for classical solvers and LR-QAOA for fully connected random problems of WMaxcut. (a) The left plot shows the TTS of SA vs. TABU search for different numbers of qubits. Each marker represents a problem. The right plot shows the TTS vs. the 100 random cases sorted by TTS of SA for 40-qubit size problems. The shaded region highlights the 20 problems with the longest TTS. (b) TTS of SA, TABU search, and CPLEX. Three configurations are used for SA and TABU in each case: 50, 100, 200, and 500 sweeps, and the lowest TTS is used for the scaling. (c) TTS of LR-QAOA for $p=50$ and $100$, the scaling is taken over the minimum TTS at each time-point. The markers represent the median value, and the error bars represent the first and third quartiles of the 20 cases distribution.}
\end{figure*}
In Figure~\ref{Fig:scaling}(a)-left, we show the TTS of SA vs. TTS of TABU search. There is a high correlation coefficient (PCC=0.7) \cite{PCC} between the solvers' TTS, indicating that random problems requiring longer TTS for one solver tend to require longer TTS for the other as well. In contrast, the PCC between TABU and CPLEX is 0.23, reflecting a weak correlation; thus, what is considered difficult for TABU is not necessarily difficult for CPLEX. Based on this information, we select the 20 cases with the longest TTS for SA out of the 100 random instances generated for each problem size. The shaded region in Fig.~\ref{Fig:scaling}(a)-right highlights the problems selected for the 40-qubit problem size.

Figure~\ref{Fig:scaling}(b) shows the scaling of SA and TABU search, and CPLEX B\&B in seconds. We use sweeps ranging from 50 to 500 for the heuristic solvers and choose the minimum TTS in each problem size case. In the case of SA, the best TTS is found using 50 to 100 sweeps. This is a consequence of the time required to implement the sweeps. While more sweeps generally lead to better solutions, the improvement does not always justify the additional evaluation time. Therefore, there is a tradeoff between the number of sweeps and the evaluation time, indicating that an optimal number of sweeps exists for a given problem size. 

The case of TABU is similar to SA; for a small number of qubits, the optimal number of sweeps is around 100. However, as the number of qubits increases, the best configurations shift. At $N_q=40$, both 200 and 500 iterations yield similar TTS. The scaling for SA is slightly better than TABU search, with a shorter TTS in all cases. At the problem sizes considered, the advantage of TABU search in maintaining a list of previously visited solutions does not appear to be necessary. As a result, the additional computational cost of comparing against this list at each iteration might affect the TTS. The TTS of CPLEX is several orders of magnitude higher than that of the other solvers, and the corresponding fitting function does not appear to be reliable.

Figure~\ref{Fig:scaling}(c) presents the TTS of LR-QAOA in arbitrary units, which must be rescaled according to the two-qubit gate time, $t_g$, of a given quantum computer. We use $p=50$ and $p=100$ and choose the best TTS from them. For visualization, we use a $t_g=2.5\times 10^{-9}$ that matches the time of TABU search at $N_q = 20$ and corresponds to a gate time of $t_g = 2.5 \ ns$. Comparing the models, LR-QAOA shows a potential scaling advantage over the other solvers. Achieving competitive scaling with LR-QAOA would likely require depths beyond $p>100$. The relative error of the perceived scaling is 0.211 for CPLEX, 0.023 for SA, 0.019 for TABU, and 0.011 for LR-QAOA.

\subsection{Experiments}

\begin{figure}
\centering
\includegraphics[width=8.5cm]{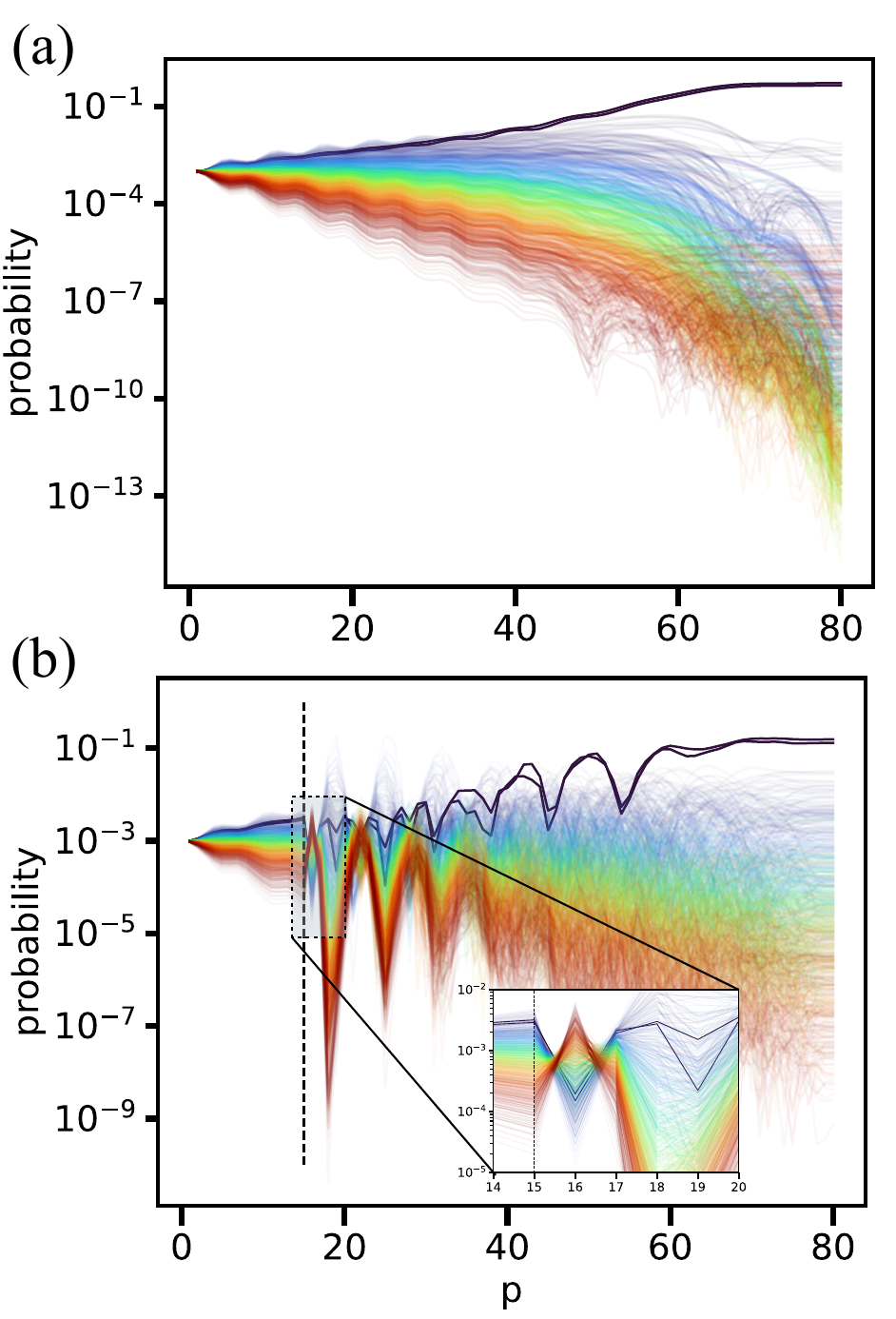}
\caption{\label{Fig:eig_evol} Probability of observing the different bitstring solutions of a 10-qubit MIS problem for the LR-QAOA protocol with p=80 (a) noiseless evolution (b) a layer of X gates applied at p=15. The lines on the graph represent various eigenvalues, with darker blue indicating lower energy and darker red indicating higher energy. The two values highlighted in dark blue correspond to the optimal solutions for the given problem.}
\end{figure}

In this section, we show numerically and experimentally how noise affects LR-QAOA. Before moving to numerical simulations of LR-QAOA under depolarizing noise, we want to show LR-QAOA's ability to overcome errors. In Fig ~\ref{Fig:eig_evol}-(a), the noiseless evolution of the eigenvalues of the cost Hamiltonian for LR-QAOA is presented. In Fig.\ref{Fig:eig_evol}-(b), the same protocol is shown but this time depicts the evolution under full inversion of the qubits using a layer of X gates applied at $p=15$. At $p=16$, the eigenvalues experience a full inversion of probabilities, with high-energy bitstrings now having a large probability. This is quickly corrected by LR-QAOA, increasing the probability of getting the optimal solution. This inversion comes with the price of a reduction in the success probability from $96.1 \%$ in (a) to $28.5\%$ in (b). Therefore, even if noisy conditions deteriorate the success probability, the errors do not completely remove the logic of the circuit.

\begin{figure*}
\centering
\includegraphics[width=17.5cm]{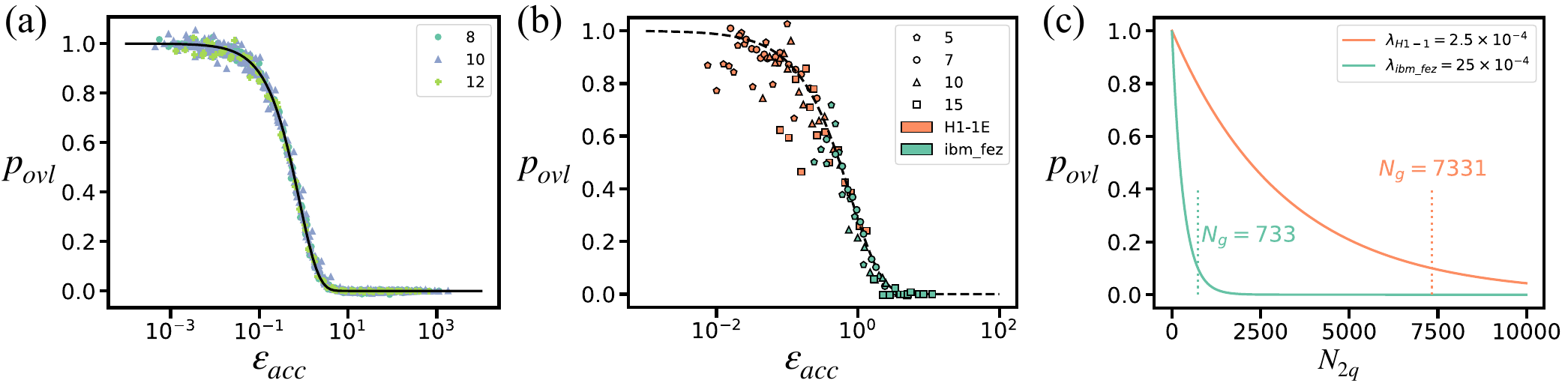}
\caption{\label{Fig:noise_model}Overlap between the success probability of LR-QAOA under noise and the ideal probability vs the accumulated error. (a) Noise simulation using depolarizing noise for 8, 10, 12-qubit problems with p=10, 20, 40, 25 depolarizing noise errors between $\lambda=10^{-5}$ and $1$, and 3 random graphs with edge density $E_d = 0.2$, 0.5, and 1. The fitting parameter in Eq.~\ref{Eq:noise} is $k_0=1.82$, which corresponds to the black line. (b) Overlap of the success probability on real QPUs, ibm\_fez (green) and Quantinuum H1-1E (orange) for 5 to 15 qubits. The errors perceived using the noise model are $\lambda_{H1-1E} = 2.5 \times 10^{-4}$ and $\lambda_{ibm\_fez} = 25 \times10^{-4}$. (c) Overlap probability vs. the number of two-qubit gates for noise models of ibm\_fez and H1-1E. The dotted line represents the number of 2-qubit gates where an overlap of 10\% is reached. }
\end{figure*}

Figure \ref{Fig:noise_model} shows simulations of different LR-QAOA configurations for the WMaxcut varying p, $\lambda$, $N_{q}$, and $E_d$ under the depolarizing noise model. We make a distinction in this figure by the number of qubits, but the markers represent cases with different p, $\lambda$, and $E_d$ as well. Therefore, even if different parameters could have an impact on the solution, the noise can be well described by the number of 2-qubit gates, the depolarizing noise, and a single fitting parameter. The fitting parameter in Eq.~\ref{Eq:noise} is $k_0=1.82$ for WMaxcut; it might be COP dependent, but further analysis is needed. 

We use this noisy model on experimental results for ibm\_fez and H1-1E. The results are shown in Fig.~\ref{Fig:noise_model}(b) for 5 to 15-qubit problems of fully connected WMaxcut. For the case of ibm\_fez, we use the parity twine chains (PTC) \cite{dreier2025connectivityaware, klaver2024} strategy to encode the LR-QAOA quantum circuit into a 1D-chain of qubits of the QPU. The number of 2-qubit gates for each layer of LR-QAOA is $N_{2q} = N_q(N_q-1)/2$ for H1-1E and $N_{2q} = N_q^2 - 1$ for ibm\_fez. Based on the noise model Eq.~\ref{Eq:depolarizing}, the fitted average error rates are $\lambda=2.5\times 10^{-4}$ for H1-1E and $\lambda=25 \times 10^{-4}$ for ibm\_fez. The  $\lambda$ can be interpreted as the average 2-qubit error of the QPU for the LR-QAOA on WMaxcut problems. These errors are similar to the average 2-qubit gate error measured by Randomized benchmarking (RB) \cite{Knill_2008}, which is $36\times 10^{-4}$ for ibm\_fez and $9\times 10^{-4}$ for H1-1.

Figure~\ref{Fig:noise_model}(c) shows the estimated overlap probability, $p_{ovl}$, as a function of the number of two-qubit gates, under the noise levels observed in ibm\_fez and H1-1E. To achieve an overlap of $p_{ovl}=0.1$ (i.e., 10\% of the ideal probability), the maximum number of two-qubit gates should be limited to approximately 733 for ibm\_fez and 7331 for H1-1E.  This limit is independent of the problem size, and therefore, can be used as an estimate of how many 2-qubit gates one can use until the overlap is too short to observe the optimal solution. Even if the error grows exponentially with the number of 2-qubit gates, the probability of success also grows exponentially as the number of layers grows. Therefore, there is a point where the trade-off between noise and LR-QAOA reaches an equilibrium point, and a maximum probability of success is obtained. A decrement in one order of magnitude in the noise leads to an increment in 1 order of magnitude in the number of 2-qubit gates that can be used, for instance, a $\lambda = 2.5 \times 10^{-5}$ and expecting an overlap of 10\% allows to use $N_g=73,310$. In the Supplementary Note 4, we extend the depolarizing noise model study to the IonQ Aria QPU.

\begin{figure*}
\centering
\includegraphics[width=17.5cm]{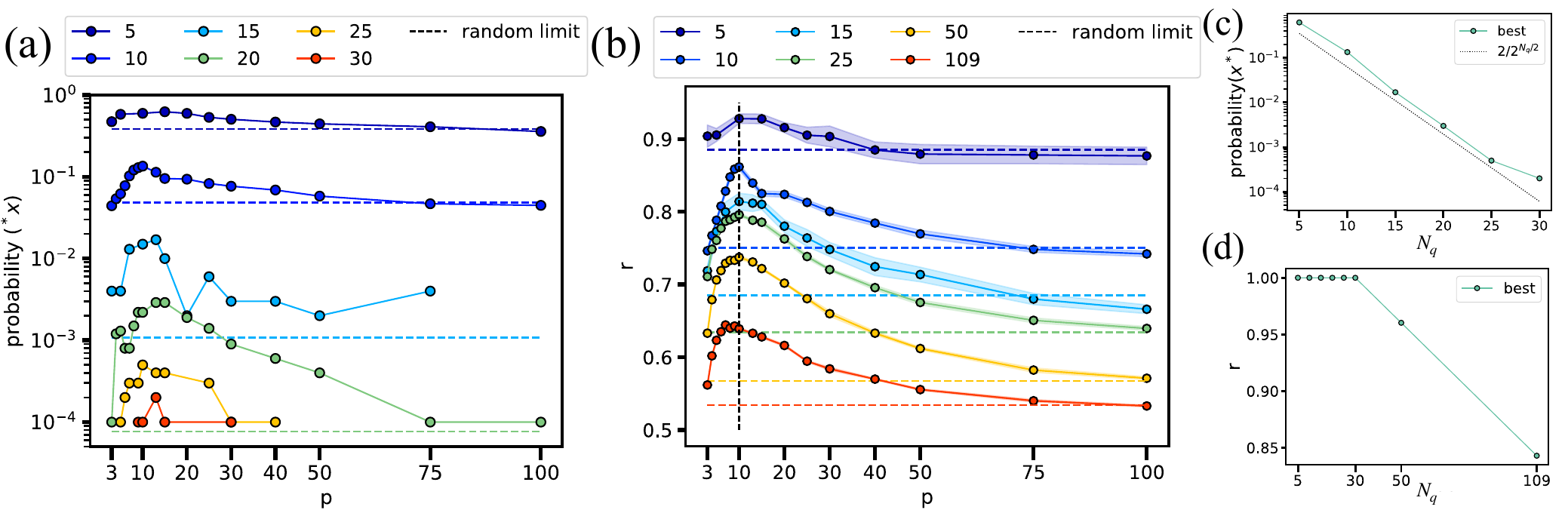}
\caption{\label{Fig:probability} Experimental results of the LR-QAOA protocol on {\it ibm\_brisbane} for WMaxcut problems. (a) Probability of success vs.~number of layers of the LR-QAOA protocol. Colors represent the number of qubits from 5 to 30 qubits. The dashed lines 'random limit' represent the success probability of a random sampler after the mitigation technique is applied for each problem with the same color. (b) Approximation ratio vs.~LR-QAOA layers. Colors represent problems from 5 to 109 qubits. The shaded region represents the standard deviation over 10 sets of 1000 shots. (c) Best probability of success of (a) vs.~number of qubits. The dashed line representing a quadratic speedup is added as a reference. (d) Best approximation ratio of all the samples in the experiment vs.~number of qubits. }
\end{figure*}

Figure ~\ref{Fig:probability}-(a) shows the probability of success vs. the number of LR-QAOA layers of random cases of the WMaxcut for variables from 5 up to 30 running on {\it ibm\_brisbane}. We use 10,000 samples for each problem size. We do not include information for larger problem sizes because no optimal solution is observed for them. The dashed line represents the probability of success of mitigated samples of a random sampler (See Sec.~\ref{A:mitig}). In other words, circles above the dashed line of its respective color cannot be explained as the result of a random process and therefore can be attributed to LR-QAOA. To contextualize our outcomes, observing the optimal solution for the 30 qubits problem with a random sampler requires, in the worst case $2^{30}/2 = 536,870,912$ evaluations of the cost function. In our experiment, we find the optimal solution 2 times at 13 layers using LR-QAOA on a noisy device using $10,000$ samples and the mitigation technique (See Sec. \ref{A:mitig} for the mitigation technique). This means $10,000 \times 30$ further evaluations, representing an improvement over random guessing of $536.870.912/310.000 \approx 1732$ times.

Figure ~\ref{Fig:probability}-(b) shows the approximation ratio of the instances of WMaxcut from 5 to 109 qubits using LR-QAOA on {\it ibm\_brisbane}. The vertical dashed line at $p=10$ indicates the number of layers for which the best performance of LR-QAOA is obtained. After $p=10$, the system is slowly moved towards a maximally mixed state. At $p=100$, it is reached in all the cases. We attribute this phenomenon to the nature of LR-QAOA, which initially improves faster than the destructive effects of noise. However, above a particular noise threshold, noise begins to dominate, leading to a monotonic decrease in the quality of the solutions obtained. This leads to an interesting behavior, for instance, at $p=3$ the approximation ratio is the same as that at $p=40$ for the 109-qubit case, despite the latter requiring roughly 13 times more time and 2-qubit gates than the former.

Figure \ref{Fig:probability}-(c) shows the maximum probability over all layers vs.~the number of qubits for the 1D WMaxcut experiment. The dashed line that represents the quadratic speedup $2/2^{N_q/2}$ over random sampling is added as a reference. In the experiments, the highest probability occurs within the range of $p=10$ to $13$. The experiments hold a similar decay to the quadratic speedup, with a shift that can be attributed to the mitigation technique. Additionally, Fig.~\ref{Fig:probability}-(d) shows the best approximation ratio among all the samples vs.~the number of qubits. The maximum average approximation ratio for the 109-qubit experiment is $r=0.64$, with the best sample having a $r=0.84$.

\begin{figure*}
\centering
\includegraphics[width=18cm]{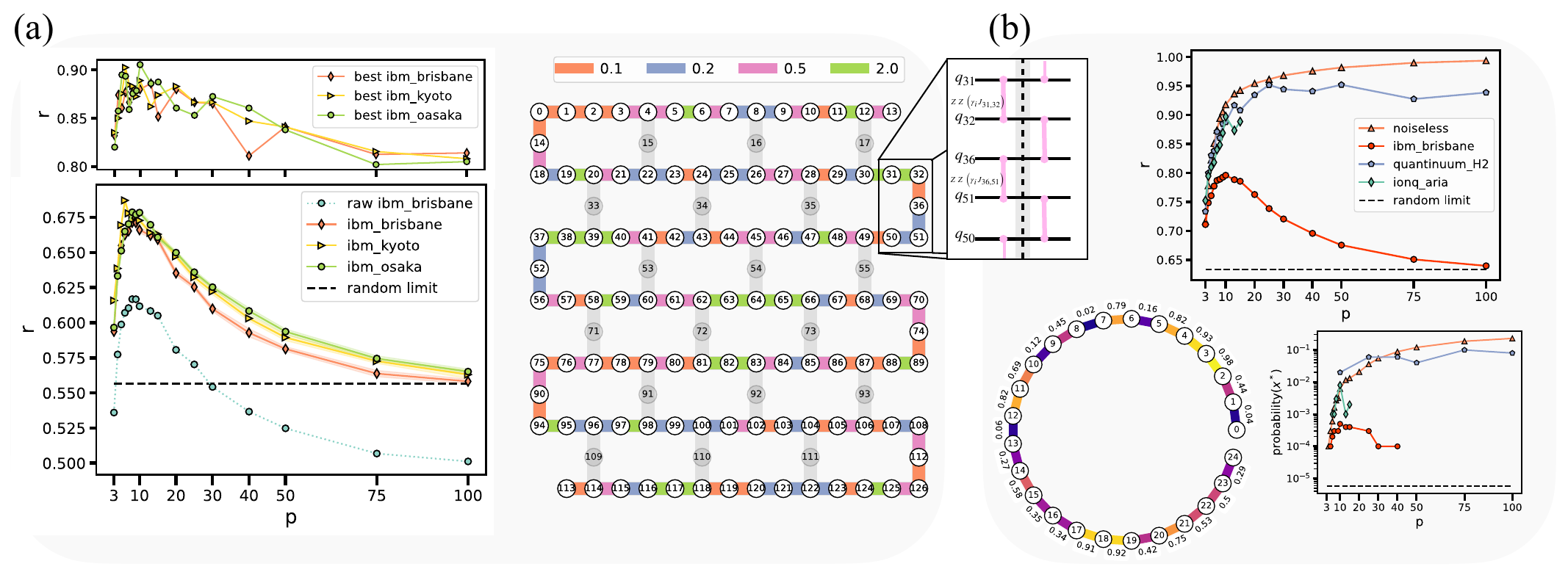}
\caption{\label{Fig:r_comparison} (a) 109-qubit WMaxcut experiment on IBM Eagle devices using LR-QAOA. The upper-left plot represents the best approximation ratio observed from the 10000 samples at each $p$. The bottom-left plot shows the average approximation ratio, where the dotted line represents the raw result from {\it ibm\_brisbane} and the solid lines represent the mitigated results over the three different IBM devices. The black dashed line is the limit where the system reaches the maximally mixed state. The right plot shows the IBM Eagle layout with the 1D random WMaxcut on it. The colors represent the random weights chosen from 4 possible values 0.1, 0.2, 0.5, and 2.0 for each edge in the graph. The inset shows how the 2-qubit gates are implemented using only a depth of 2 for each LR-QAOA layer. The shaded region represents the standard deviation over 10 sets of 1000 shots. (b) 25-qubit WMaxcut experiment comparative analysis for 3 different vendors. The bottom-left plot presents the graph of the WMaxcut selected with the corresponding weight values. In the upper plot, the approximation ratio of the devices with the triangles, circles, diamonds, and pentagons corresponds to the mitigated results from a noiseless simulation, {\it ibm\_brisbane}, {\it ionq\_aria}, and {\it quantinuum\_H2}, respectively. Because of a limitation in the maximum number of single-qubit and two-qubit gates, there are no results above $p=15$ using {\it ionq\_aria}. In the bottom right plot, the probability of success for the same experiment is shown.}
\end{figure*}

Figure ~\ref{Fig:r_comparison}-(a) shows the approximation ratio of the 109 qubits WMaxcut problem using LR-QAOA from $p=3$ to $p=100$. At $p=10$, the maximum approximation ratio is reached for the three devices \textit{ibm\_brisbane}, \textit{ibm\_kyoto}, and \textit{ibm\_osaka}. The noise at larger $p$ leads the system towards a maximally mixed state, so we include the dashed line that represents the approximation ratio $r = 0.5326$ of a random sampler after the mitigation technique is applied. Unexpectedly, at $p=100$, results for \textit{ibm\_kyoto} and {\it ibm\_osaka} still deviate from the random sampler and therefore some information of the LR-QAOA protocol is present. The circuit used requires 21200 CNOT gates and a total time of $\approx 132 \mu s$. This is an indication of the resilience of the LR-QAOA to noise. 

At this scale, we surpass the point where exact classical simulation of LR-QAOA is feasible both in terms of the number of qubits and depth of the circuit. The 1D connectivity of the graph makes the simulation of the LR-QAOA suited for approximation methods based on tensor networks.  If the problem connectivity is increased, making it hard to be simulated classically, this experiment might be presented as a quantum utility experiment. This means that a classical algorithm cannot mimic the sampling properties of LR-QAOA at large $p$ and $N_q$. There are different techniques proposed for addressing the simulation of quantum supremacy \cite{Arute2019} or utility \cite{Kim2023} experiments after their publication (e.g., \cite{Pednault2017, Begusic2024}), so the validation of this remains subject to evaluation within the research community. Independent of the answer, these results indirectly imply the efficacy of LR-QAOA to solve COPs in scenarios involving more than 42 qubits.

Figure ~\ref{Fig:r_comparison}-(b) presents a comparative analysis between {\it ionq\_aria}, {\it quantinuum\_H2}, and {\it ibm\_brisbane} in solving a 25-qubit instance of the WMaxcut problem. The number of samples is 10000 for the noiseless simulator and {\it ibm\_brisbane}, 1000 for {\it ionq\_aria}, and 50 for {\it quantinuum\_H2}. The performance of {\it quantinuum\_H2} stands out, achieving a maximum approximation ratio of $r = 0.95$ at $p=50$, compared to $r=0.98$ at $p=50$ of the noiseless simulator, {\it ibm\_brisbane}'s $r = 0.80$ at $p=10$, and {\it ionq\_aria} $r = 0.90$ at $p=10$.

From a time perspective, executing an instance of WMaxcut LR-QAOA for $p=10$ on {\it ibm\_brisbane} requires approximately $\approx 13.2 \mu s$, whereas {\it ionq\_aria} completes the same task in about $\approx 144 ms$, and {\it quantinuum\_H2} in $\approx 36 ms$. The three devices successfully identify the optimal solution for this problem, with {\it quantinuum\_H2} achieving a maximum probability of success of 0.10 at $p=75$, {\it ionq\_aria} achieving a maximum probability of success of 0.008 at $p=10$, and {\it ibm\_brisbane} reaching 0.0005 at $p=10$. This means that {\it quantinuum\_H2} is 12.5 times more effective in finding the optimal solution than {\it ionq\_aria}, and 200 than {\it ibm\_brisbane}. 

However, the accuracy gain for {\it quantinuum\_H2} does not fully compensate for the time required for sampling. In other words, for every optimal sample obtained from {\it quantinuum\_H2}, one could obtain approximately 2700 samples on {\it ibm\_brisbane}. To observe an optimal solution at $p=10$ using {\it ibm\_brisbane} we need $\approx 13.2\times10^{-6}/0.0005 =0.0264 s$ while {\it quantinuum\_H2} $\approx 36\times10^{-3}/0.02 = 1.8 s$. This means when {\it quantinuum\_H2} finds a solution, {\it ibm\_brisbane} has already found 68.

\section{Discussion}\label{Sec:Conclusions}

In this work, we have presented numerical and experimental evidence that LR-QAOA constitutes an effective QAOA protocol. This means that this protocol works efficiently for the problem tested, increasing the probability of success as the number of layers increases. We simulate MIS, BPP, TSP, Maxcut, WMaxcut, 3-Maxcut, KP, PO, Max-2-SAT, and Max-3-SAT problems with up to 42 qubits and 400 layers on the modular supercomputer JUWELS. Additionally, we test LR-QAOA using WMaxcut problems from 5 to 109-qubit cases and $p$ from 3 to 100 on real quantum hardware using {\it ibm\_brisbane}, {\it ibm\_osaka}, {\it ibm\_kyoto}, {\it quantinuum\_H2}, and {\it ionq\_aria} finding that LR-QAOA is resilient to noise. We show that this behavior arises from the algorithm's ability to enhance solution quality at a rate that initially outpaces the accumulation of noise. While the overlap of the probability of success decreases exponentially with the number of 2-qubit gates, it is compensated by an exponential growth in the probability of success up to some p. This explains why the highest probability of success does not show up at the smallest number LR-QAOA layers in the experiments but at some other point, e.g., at p=10 in the ibm\_brisbane case.

One important conclusion from this work is that one can completely suppress the classical optimization step in QAOA for some COPs. With the fixed schedule in LR-QAOA, one can reduce the set of parameters to tune to only three $\Delta_\beta$, $\Delta_\gamma$, and $p$.

We show the evolution of LR-QAOA from the perspective of the amplitudes of the computational basis states. This change in framework allows us to explain the evolution of the amplitudes under the application of $U_C$ and $U_B$. Under the application of $U_C$, each amplitude is rotated proportionally to the state energy. The case of $U_B$ is more complex, but every amplitude evolves with contributions from the other states' amplitudes with the Hamming distance as the indicator of how to group their contribution. The annealing characteristics of LR-QAOA, along with a constant rotation of the states' amplitude (constant slope of the linear ramp) under $U_C$, allow the exploitation of an interference pattern that enhances the optimal solution in the different COPs.

We observe that the success probability of the optimal solutions using LR-QAOA for the different COPs seems to scale as $probability(x^*) \approx 2^{-\eta(p) N_q + C}$ for $\eta(p)$ decreasing with $p$ and a constant $C$. We add further evidence to solve fully connected WMaxcut problems. We create 100 random problems and select the 20 of them that require the highest number of iterations for the SA solver. We compare SA, LR-QAOA, TABU, and CPLEX in terms of TTS for these problems, observing a better scaling in LR-QAOA.

We extend the study to Maxcut, Max-2-SAT, and Max-3-SAT with up to 42 qubits. Using $p=N_q$, we find that on average, the $ probability (x^*)$ remains nearly constant for WMaxcut, Maxcut, and Max-3-SAT. The Max-2-SAT case is an exception, using $p=N_q$, it shows an exponential decay in the probability of success, still above a quadratic speedup over random guessing. We think this is a consequence of problems with a high concentration of solutions close to the optimal solution.

Moreover, we find that LR-QAOA tolerates noise. This is important as we are at a stage where quantum computers have moderate noise. We simulate a MIS using LR-QAOA with $p=80$, and at $p=15$ we add a layer of X gates to see how the algorithm evolves under this noise. We find that even in this scenario, the error does not expand, and the optimal solution can still be found with high probability. We extend the study of noise using depolarizing noise on an FC WMaxcut with different numbers of qubits, layers, and $E_d$, and find that $p_{ovl}$ decreases exponentially with the number of 2-qubit gates. Using the same model, we fit experimental results in a Quantinuum H1-1 emulator and ibm\_fez QPU. In both cases, our model fit well with an apparent error of $\lambda = 2.5 \times 10^{-4}$ for H1-1E and  $\lambda = 25 \times 10^{-4}$ for ibm\_fez. These errors allow the execution of 733 gates in the case of ibm\_fez and 7331 for H1-1E to have an overlap of 10\% with the ideal probability. 

We find that there is an effective number of layers for which the real device shows the best performance. We call it the $p_{\mathrm{eff}}$, this parameter can be used to measure the progress of quantum technology for combinatorial optimization. For IBM Eagle devices and {\it ionq\_aria} $p_{\mathrm{eff}}=10$ and for {\it quantinuum\_H2} is $p_{\mathrm{eff}}=50$ for a 1D topology problem. We expect the $p_{\mathrm{eff}}$ decreases for a fully connected graph problem because the number of two-qubit gates per layer grows by $O(N_q^2)$ compared with the 1D case, $O(N_q)$.

The experimental results make us optimistic that LR-QAOA can keep high performance even in the presence of noise. For example, {\it quantinuum\_H2} already shows its peak performance at $p=50$ and loses little performance at $p=100$. At the peak point, the device reaches the best approximation ratio of $r=0.95$ and $probability(x^*) = 0.08$ for a 25-qubit problem. On the other hand, the inaccuracy of {\it ibm\_brisbane} is still compensated by its sample rate for the same problem.

In a recent study \cite{koch2025quantumoptimizationbenchmarklibrary}, 10 hard COPs for classical solvers were introduced, many of which are sparsely connected; this characteristic can make them suitable to be tackled by LR-QAOA. Between these problems is the MIS, and hard instances show up at sizes with a few hundred qubits. For instance, there is a case with 500 qubits and 6256 edges (See Table 7. R 500 005 1 in \cite{koch2025quantumoptimizationbenchmarklibrary}) for which the optimal solution is not known. For a p=200 LR-QAOA and based in our perceived scale, we need 1'250,200 2-qubit gates, the noise in this case to reach an overlap of 10\% is $\lambda=1.45\times 10^{-6}$ and based on the scaling of Fig. \ref{Fig:probabilities} the number of samples needed is around 25,000 to observed the optimal solution. Currently, the noise in the QPUs is 2 orders of magnitude above the level of error needed, and the number of qubits is at most 156. 

It might be possible that the noise level required of a fault-tolerant quantum computer (FTQC). A recent effort to estimate the overhead scenario of FTQC has been presented for the 8-SAT problem \cite{omanakuttan2025threshold}, finding that at some point, QAOA combined with amplitude amplification with the FTQC overhead can still outperform the best classical solver for that problem.

\section*{Data Availability}
All study data are included in this article and the Supplementary Materials. The datasets for the problems used and/or analyzed during the current study are available from the following publicly accessible repository \url{https://github.com/alejomonbar/LR-QAOA}.

\section*{Code availability}
The code is available from the following publicly accessible repository \url{https://github.com/alejomonbar/LR-QAOA}.

\begin{acknowledgments}
\vspace{-10pt}

The authors thank Dennis Willsch, Vrinda Mehta, Hans De Raedt,  and Fengping Jin for the insightful discussions and suggestions made for the present work. J. A. Montanez-Barrera acknowledges support from the German Federal Ministry of Education and Research (BMBF), funding program Quantum Technologies - from basic research to market, project QSolid (Grant No. 13N16149).
The authors gratefully acknowledge the Gauss Centre for Supercomputing e.V. (www.gauss-centre.eu) for funding this project by providing computing time on the GCS Supercomputer JUWELS at Jülich Supercomputing Centre (JSC).

This research used resources of the Oak Ridge Leadership Computing Facility for the experiments on {\it quantinuum\_H2}, which is a DOE Office of Science User Facility supported under Contract DE-AC05-00OR22725.

\end{acknowledgments}

\clearpage  
\bibliography{References}
\bibliographystyle{apsrev4-2}

\clearpage  
\onecolumngrid  

\appendix

\section*{Appendix}
In these Supplemental Notes, we provide detailed insights into the LR-QAOA algorithm's parameter selection, performance behavior under noise, and its application across various combinatorial optimization problems. In Sec.~\ref{Sec:2D-PD}, we analyze the parameter selection landscape of LR-QAOA. We explore how the values of $\Delta_\gamma$ and $\Delta_\beta$ influence the success probability across different COPs at a fixed depth of $p=50$, and highlight the presence of a performance ridge consistent across normalized Hamiltonians. In Sec.~\ref{A:eta}, we examine the scaling coefficient $\eta(p)$ by fitting success probabilities to the model presented in the main text. We show that for most problems, the choice of minimum qubit count $N_{q_{\text{min}}}$ has little effect on $\eta(p)$, validating the robustness of the scaling expression used. In Sec.~\ref{A:Simulations}, we present additional simulation results using fixed values of $\Delta_\gamma = 0.6$ and $\Delta_\beta = 0.3$. These results use problems with up to 42 qubits and a wide range of $p$ values, demonstrating significant amplification in success probability with increasing layers of LR-QAOA. In Sec.~\ref{A:noise-aria}, we compare the performance of LR-QAOA under depolarizing noise with real quantum hardware results from the {\it ionq\_aria} backend from IonQ. We show that the experimental behavior aligns closely with simulations at a noise strength of $\lambda = 6 \times 10^{-4}$, and discuss how the algorithm’s robustness varies with $\lambda$. In Sec.~\ref{A:UB_explained}, we present a derivation of the equations presented in Sec. \ref{Sec:Properties-LR-QAOA}, providing the mathematical formulation of the role of the unitaries involved in LR-QAOA dynamics. In Sec.~\ref{Max-3-SAT}, we introduce the Max-3-SAT problem as a representative Boolean satisfiability problem with clauses, and discuss its encoding within the QAOA framework. In Sec.\ref{sec:performance-diagram}, we review the concept of the performance diagram as introduced in \cite{Kremenetski2021}, and show how it provides a clear visualization of LR-QAOA performance. We illustrate how protocol efficiency can vary with $\Delta$ and $p$, and demonstrate equivalence between deep and shallow LR-QAOA regimes for certain success probabilities.

\subsection{Choosing $\Delta_{\gamma}$ and $\Delta_{\beta}$ parameters}\label{Sec:2D-PD}
\begin{figure*}
\centering
\includegraphics[width=17cm]{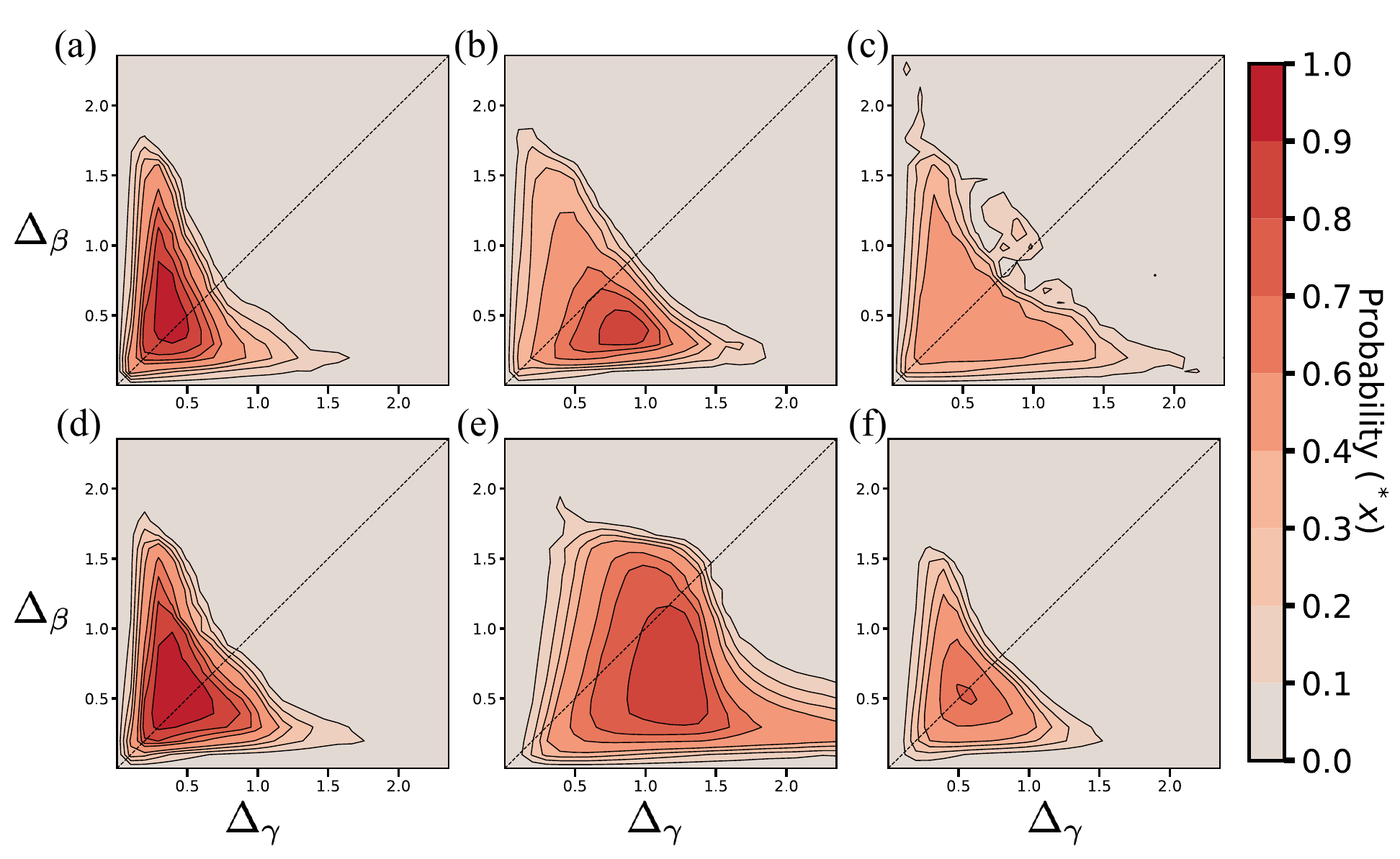}
\caption{\label{Fig:landscapess} 2D Performance diagram of $\Delta_\beta$ vs.~$\Delta_\gamma$ for the LR-QAOA protocol with p=50 for a random instance of (a) 12-qubit MIS, (b) 12-qubit KP, (c) 9-qubit TSP, (d) 12-qubit PO, (e) 12-qubit BPP, and (f) a 16-qubit WMaxcut.}
\end{figure*}

Figure~\ref{Fig:landscapess} shows the success probability landscape of LR-QAOA for values of $\Delta_\beta$ and $\Delta_\gamma$ ranging from 0 to $3\pi/4$ at $p = 50$ across different COPs. Surprisingly, consistent behavior is observed across the different plots, although the structure of the Hamiltonian varies considerably between them. As discussed in Sec. \ref{SubSec:H}, such consistency is only observed when the Hamiltonian is normalized. The red region in these plots resembles the ridge region, i.e., a region where QAOA has the highest performance in the performance diagram (See \cite{Kremenetski2023}), indicating that excessively large values of $\Delta_\beta$ or $\Delta_\gamma$ do not lead to improvement. Moreover, the best performance does not necessarily occur when $\Delta_\beta=\Delta_\gamma$, suggesting potential performance improvement through parameter tuning. In our case, a $\Delta_\beta \leq 0.6$ and a $\Delta_\gamma \leq 0.6$ give favorable results.

\begin{figure*} 
\centering
\includegraphics[width=17cm]{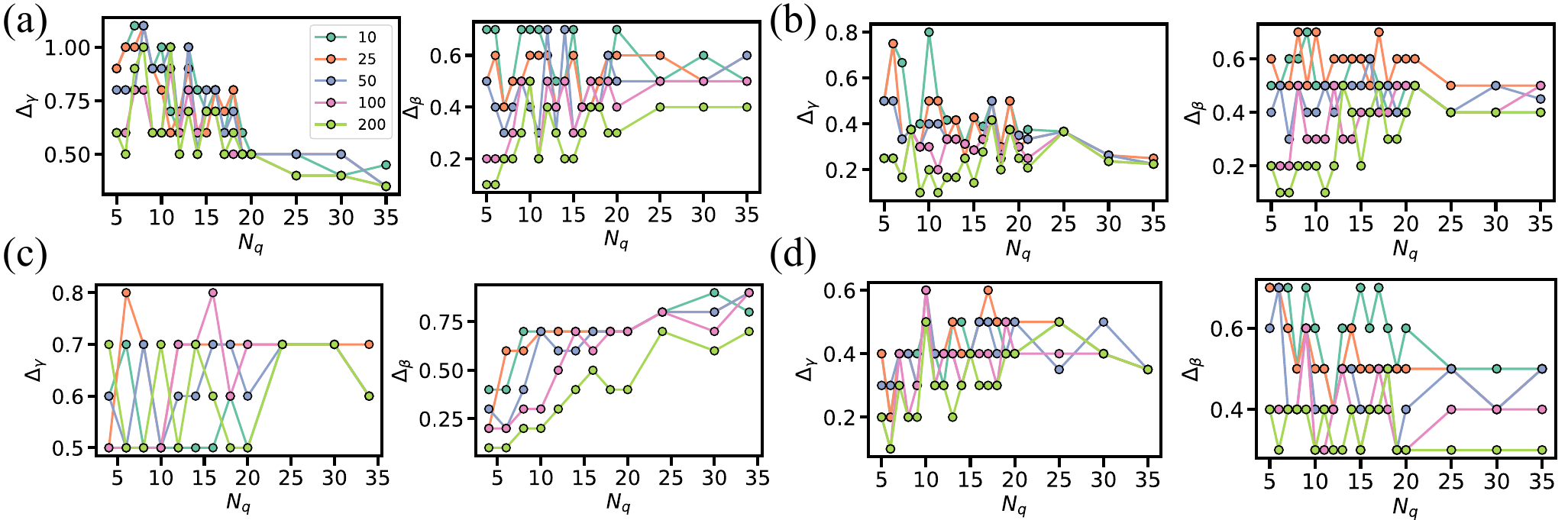}
\caption{\label{Fig:deltas}$\Delta_\gamma$ and $\Delta_\beta$ values vs. number of qubits used for the 100 instances of (a) WMaxcut, (b) MIS, (c) 3Maxcut, and (d) PO. The colors represent the value used for a given $p$ layers of LR-QAOA and the legend is given in (a).}
\end{figure*}

Figure~\ref{Fig:deltas} shows the values of $\Delta_\gamma$ and $\Delta_\beta$ used for the problems presented in Fig.~\ref{Fig:probabilities} of the main text using the strategy of scanning the $\Delta$ values for one problem size instance and reusing it on the other instance. This strategy is not ideal, but because of computational limitations to simulate large problem sizes, and the number of cases used led we to adopt this approach. The results certainly can be improved by tuning the parameters for each instance independently, but even with this approximation, the results are promising. For problems with 20 or fewer qubits, we scan over 6 $\Delta_\beta$ values and 7 $\Delta_\gamma$ values, and for more than 20, 4 $\Delta_\beta$ and 4 $\Delta_\gamma$. The parameters tend to concentrate in specific regions as the number of qubits grows, always with a tendency that $\Delta_\gamma$ and $\Delta_\beta$ for larger $p$ are less than or equal to values for smaller $p$, but a deeper study in this respect out of the scope of the present work is needed to conclude this characteristic.

\begin{figure*} 
\centering
\includegraphics[width=15cm]{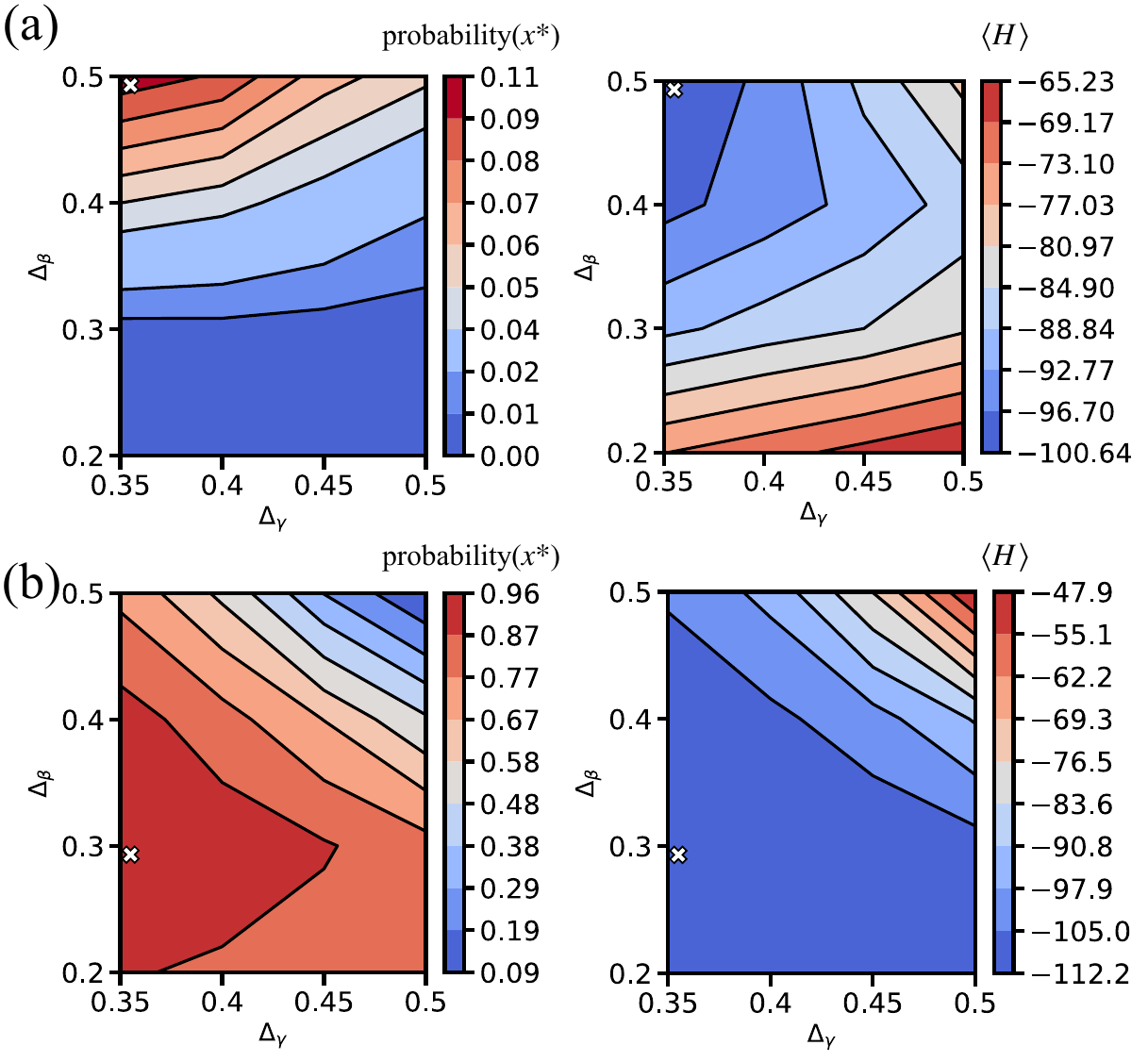}
\caption{\label{Fig:landscape_PO} Scanning of the PO for $N_q=35$ $\Delta_\gamma$ and $\Delta_\beta$ parameters (a) p=10 and (b) p = 200. The left plot is in terms of the probability of success, and the right is the average energy. The X marker is the value used in the optimization.}
\end{figure*}

Figure~\ref{Fig:landscape_PO} shows the scan of the PO first instance for $N_q = 35$. In this case, as the number of layers increases, the landscape changes considerably, and the point where the best probability is obtained also changes.

\subsection{Scaling coefficient $\eta(p)$}\label{A:eta}

\begin{figure*}
\centering
\includegraphics[width=17cm]{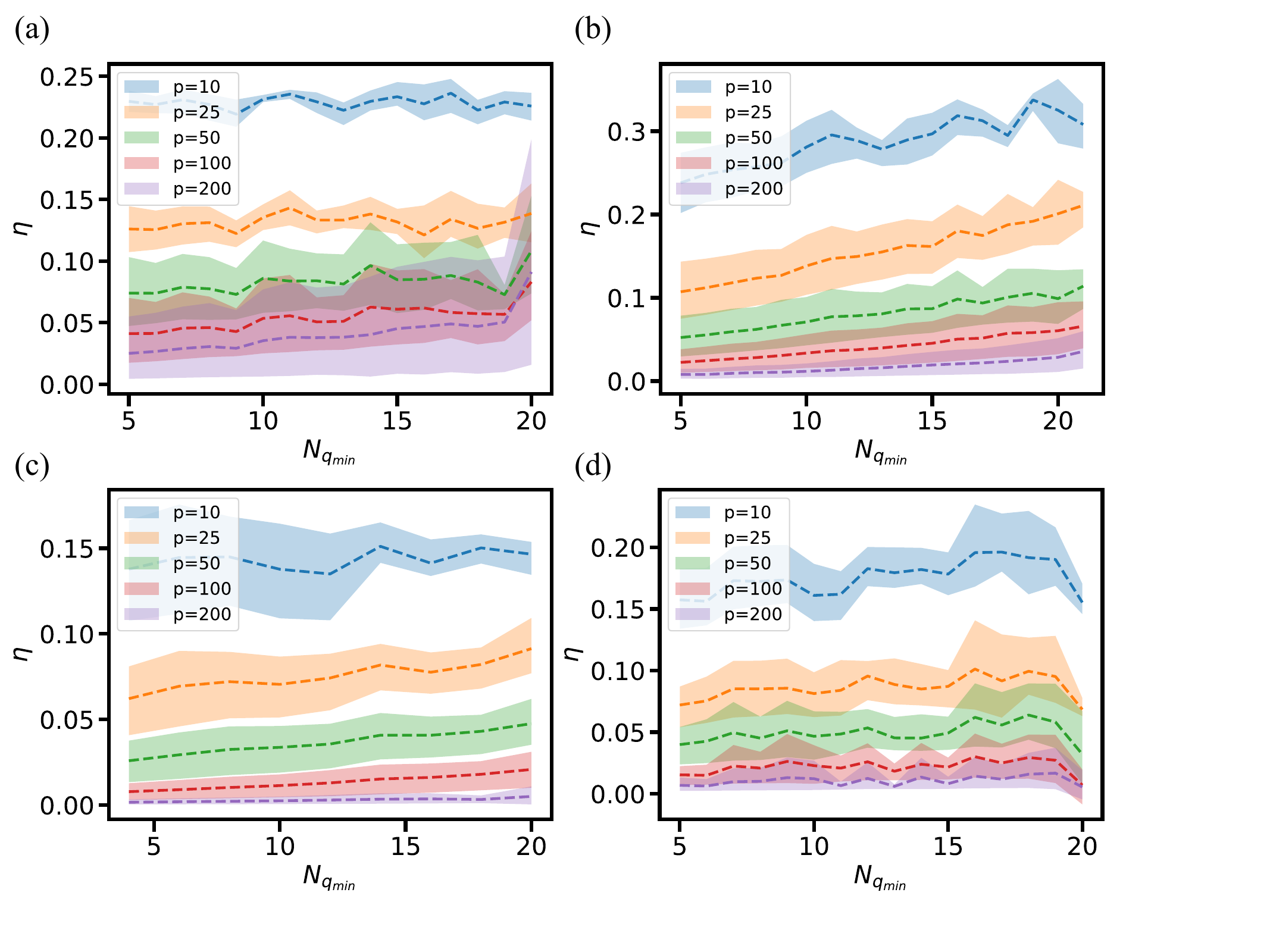}
\caption{\label{Fig:eta}$\eta(p)$ vs. minimum qubit number, $N_{q_{min}}$, used for the cutoff of the fit for (a) WMaxcut, (b) MIS, (c) 3Maxcut, (d) PO. The dashed line corresponds to the mean value and the shaded region corresponds to the quartiles Q1 and Q3 of the 100 instances.}
\end{figure*}

In Fig. \ref{Fig:eta}, the $\eta(p)$ values are shown when the minimum number of qubits used for fitting is $N_{q_{min}}$ of Eq.\ref{Eq:eta} of the main text. In all cases except MIS, the choice of $N_{q_{min}}$ has minimum impact on $\eta(p)$, supporting that Eq. \ref{Eq:eta} of the main text effectively describes the probability of success. However, it is important to note that the number of qubits ($\le 35$) is too small to draw definitive conclusions.

In the case of relative error of the probability of success vs. the minimum number of qubits used for the cutoff of the fitting function, it is observed that in the case of WMaxcut, 3Maxcut, and PO, there is no big difference if the minimum number of qubits used is 5 or 20 qubits. However, in the case of MIS seems that the scaling is changing and taking small values of $N_{q_{min}}$ affects the error observed mainly for p=10 and 25. We decided to use $N_{q_{min}} = 10$ in Fig.\ref{Fig:probabilities} of the main text.

\begin{figure*}
\centering
\includegraphics[width=17cm]{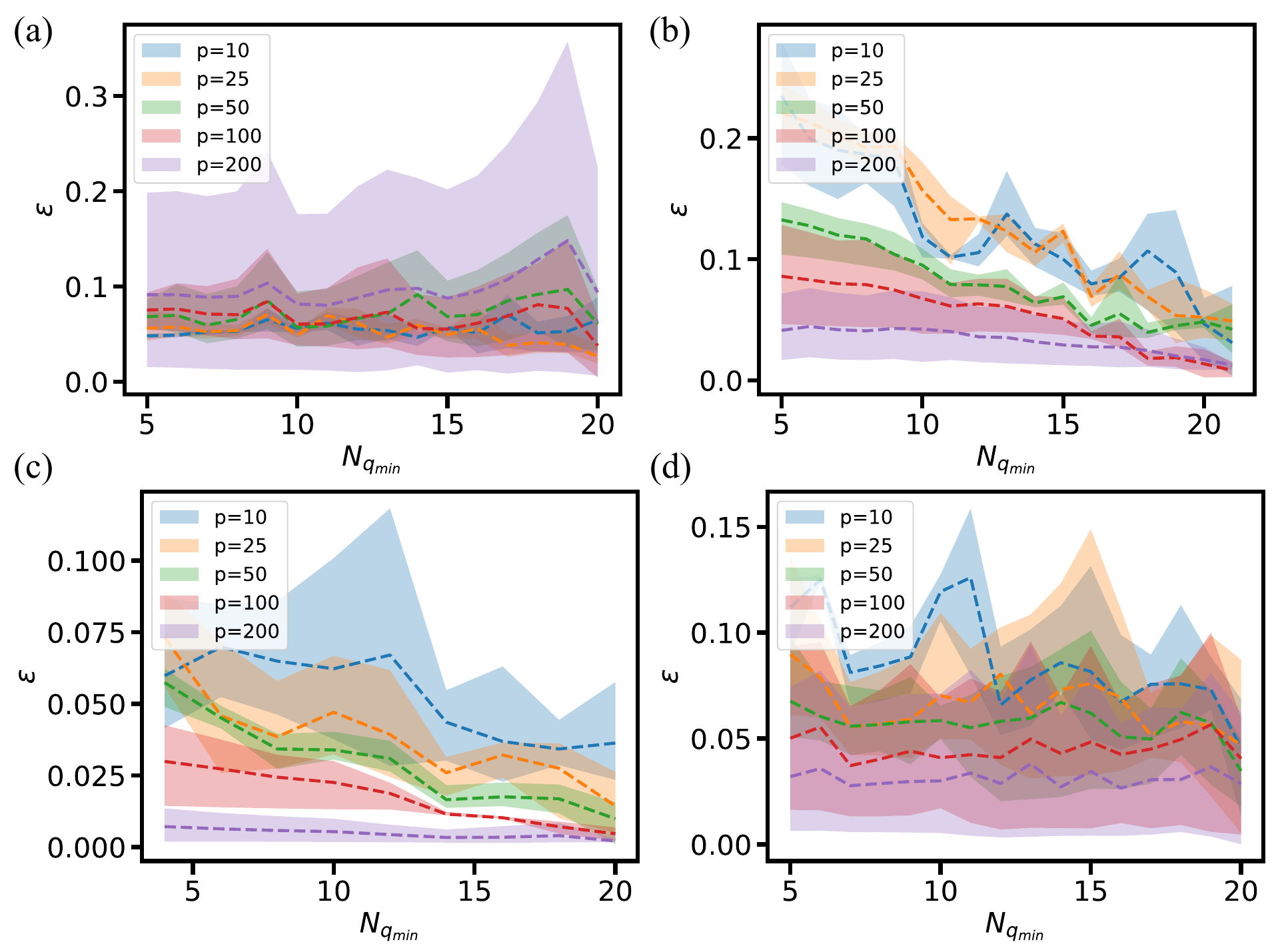}
\caption{\label{Fig:fit_etas_error} Relative error vs. minimum qubit number used for the cutoff of the fit for (a) WMaxcut, (b) MIS, (c) 3Maxcut, (d) PO. The dashed line corresponds to the mean value, and the shaded region corresponds to the fit for the quartiles Q1 and Q3 of the 100 instances.}
\end{figure*}

\subsection{Simulations using constant $\Delta_{\beta}$ and $\Delta_\gamma$}\label{A:Simulations}
\begin{figure*}
\centering
\includegraphics[width=17cm]{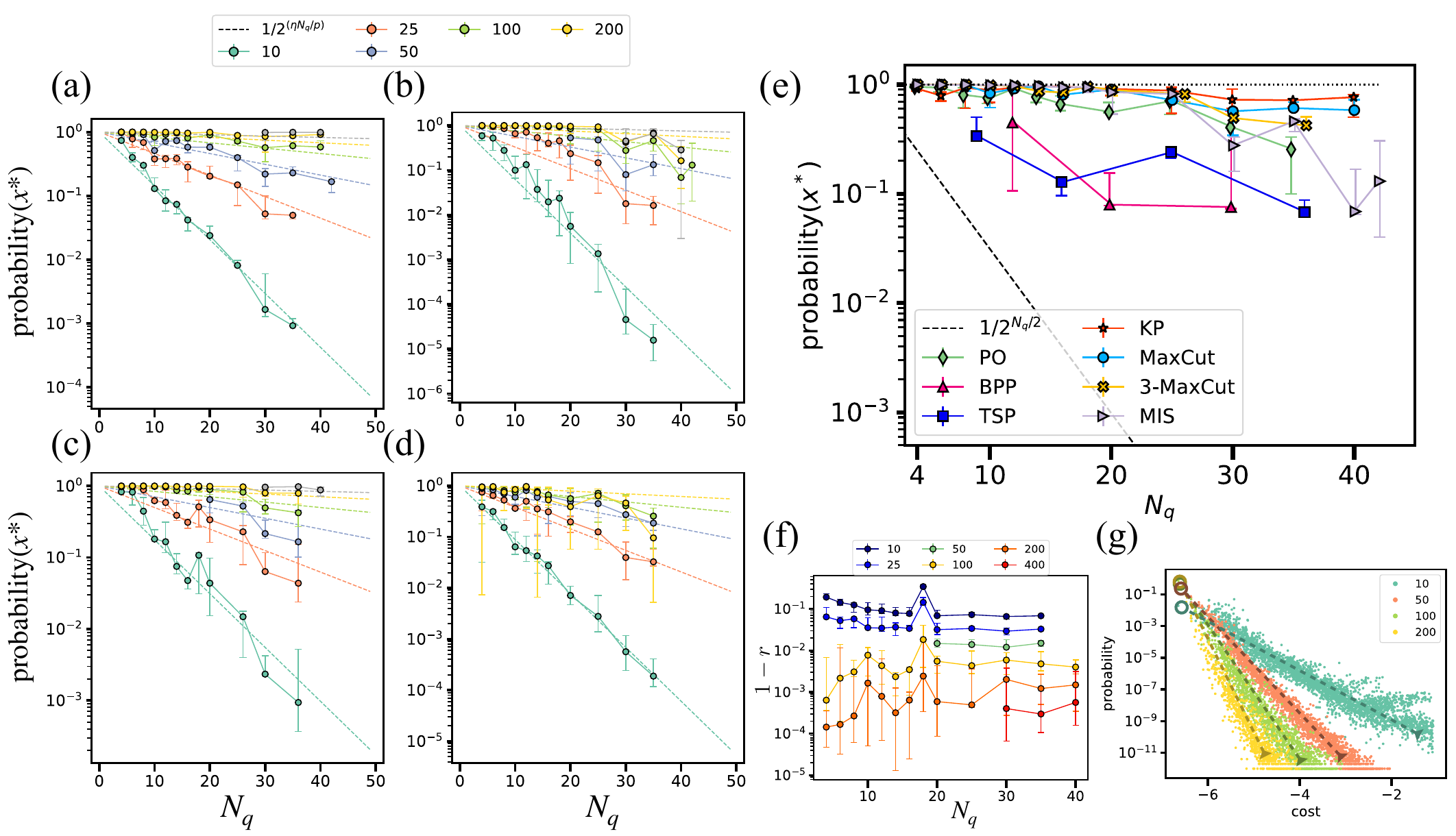}
\caption{\label{Fig:probabilitiesB} Probability of success for 5 random instances of (a) W-MaxCut, (b) MIS, (c) 3-MaxCut, and (d) PO. The error bars represent the minimum and maximum over the 5 cases. The different colors represent the number of LR-QAOA layers (see legend). Dashed lines represent the conjectured scaling $2^{-\eta N_q/p}$ for each $p$. (e) Probability of success for 5 random instances for different COPs using a $p=100$ LR-QAOA. Markers represent the median value, and error bars represent the Q1 and Q3 quartiles. The average performance of LR-QAOA as measured by (f) the fractional error, $1 - r$, of the instances of W-MaxCut from (a), and (g) probability vs.~cost of an instance of a 12-qubit PO solved using LR-QAOA with different $p$ from 10 to 200.}
\end{figure*}

Additional to the 100 cases shown in the main text, Fig.~\ref{Fig:probabilitiesB}-(a)-(d) show the probability of success for different COPs vs.~the number of qubits. We test problems with 4 to 42 qubits using the LR-QAOA with $p=10$ to $400$, $\Delta_\gamma = 0.6$, and $\Delta_\beta = 0.3$. The error bars represent the minimum and maximum $probability(x^*)$ over 5 random cases, and the circles are the mean value. The colors represent the number of layers $p$ used. Figure~\ref{Fig:linear_ramp_schedule}-(d) of the main text shows one of the 5 random cases of W-MaxCut with 42-qubits of Fig.~\ref{Fig:probabilitiesB}-(a). In this example, just employing 50 layers of LR-QAOA results in an amplification of 12 orders of magnitude in the probability of success from approximately $2/2^{42}\approx 4.54 \times 10^{-13}$ to $0.32$. To provide context, if we were to solve the same task using Grover's algorithm \cite{Grover1996}, it would require approximately $\sqrt{2^{42}}/2\approx 2\times 10^6$ iterations of the oracle, along with the diffuser. The dashed lines in Figs.~\ref{Fig:probabilitiesB}-(a)-(d) are added as guiding lines. They represent $2^{-\eta N_q/p}$ where $\eta$ is a constant. The $\eta$ values used are (a) $\eta=2.8$, (b) $\eta=4$, (c) $\eta=2.5$, and (d) $\eta = 3.5$. These results suggest that the $probability(x^*) = 1/2^{(\eta N_q/p)}$ for the random instances of different COPs analyzed. But after a careful evaluation of the scaling using a larger dataset, we find that the model that best describes the probability of success is $probability(x^*) =2^{-\eta(p) N_q + C}$ with $\eta(p)$ decreasing rapidly as $p$ grows. 

Figure~\ref{Fig:probabilitiesB}-(e) shows the probability of success for different combinatorial optimization problems up to 42 qubits using the LR-QAOA protocol with $p=100$. The dashed line, $1/2^{N_q/2}$, is added as a guiding line; it reflects a quadratic speedup in the search space. The best performance is obtained for W-MaxCut with an average $probability(x^*) = 0.58$ at 40 qubits which, compared to the optimal solution initial amplitude  $2/2^{40}\approx 0.18 \times 10^{-11}$ is an 11 orders of magnitude increment in only 100 layers of LR-QAOA. The lowest performance is for the TSP and BPP; for example, for a 6 cities problem (36 qubits) we get an average $probability(x^*) = 0.15$ which compared with the random guessing probability $1/2^{36} \approx 0.145 \times 10^{-10}$ is a gain of 10 orders of magnitude. However, it is worth noting that the number of feasible solutions to this problem is $(cities-1)!/2 = 60$, resulting in a random guessing probability within the feasible space of approximately $1/60 \approx 0.0166$.

Figure~\ref{Fig:probabilitiesB}-(f) shows the average performance of LR-QAOA for the W-MaxCut problem in terms of the fractional error $1 - r$ vs.~the number of qubits, where $r$ is the approximation ratio given by Eq.~(\ref{eq:r}) of the main text. The maximum number of qubits used is 40. Different colors represent the number of layers from 10 to 400. As expected, an increase in the number of layers leads to an increase in the average performance of LR-QAOA to find better solutions. Additionally, the relation of $1-r$ tends to stay constant with an increment in the number of qubits; this aligns with the minimum performance guarantee of MaxCut for a fixed number of QAOA layers \cite{Wurtz2021}.

Figure~\ref{Fig:probabilitiesB}-(g) shows a 12-qubit PO probability vs.~the cost of the first 3000 sorted eigenvalues using LR-QAOA from 10 to 200 layers. The dots represent the probability of getting a certain cost Hamiltonian eigenvalue vs.~the cost associated with it. This plot shows another characteristic of LR-QAOA, namely that the probability of obtaining a given solution drops exponentially with increasing cost. Even if the optimal solution is not observed in one sample of LR-QAOA, it is more likely that a low-cost energy is observed.

\begin{figure*}
\centering
\includegraphics[width=18cm]{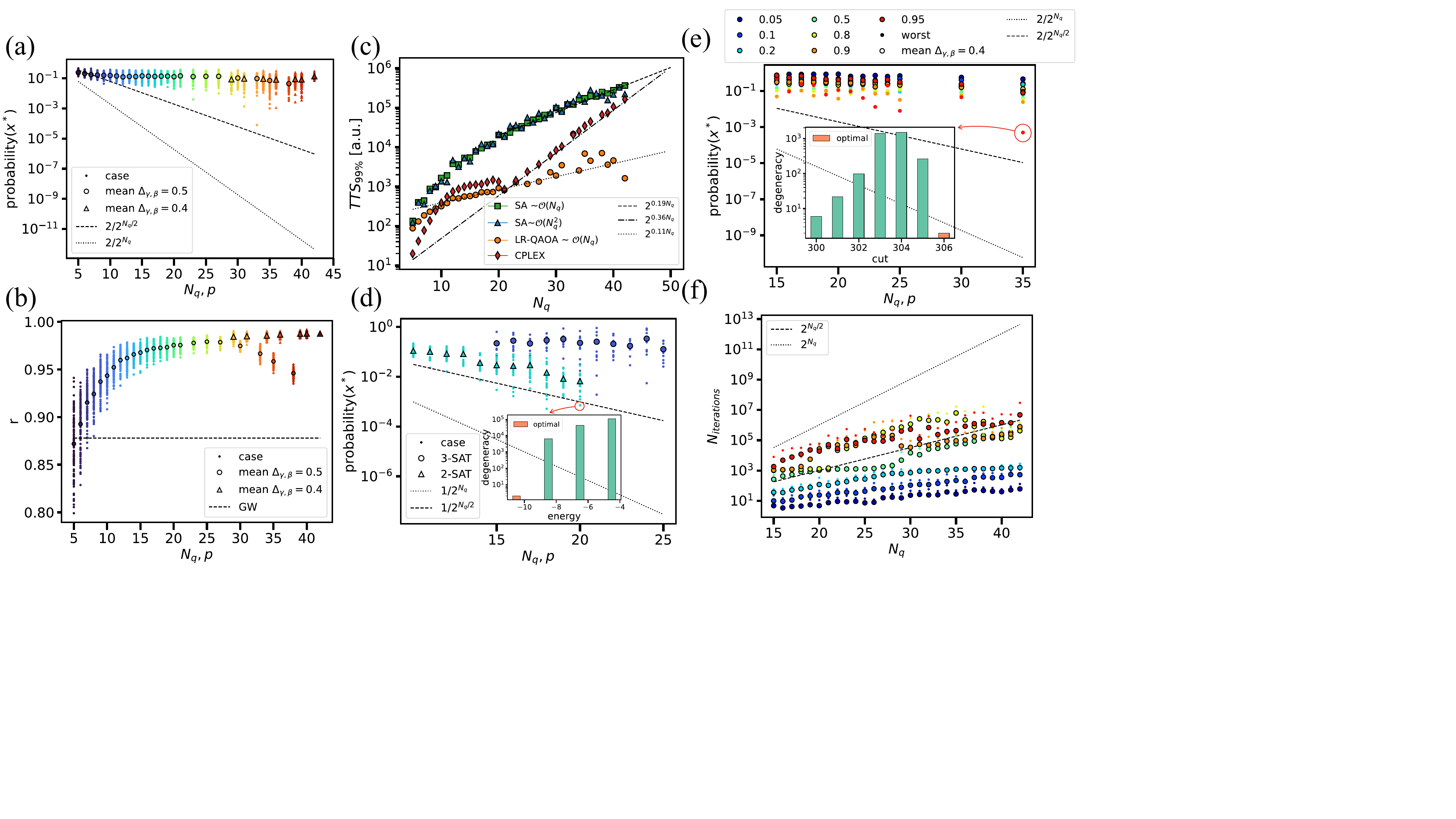}
\caption{\label{Fig:conjecture} The problems shown here are solved using LR-QAOA with $p=N_q$ (a) Probability of success of fully connected W-MaxCut problem using 100 random W-MaxCut problems up to 40 qubits and 10 problems for 42 qubits. The different vertically stacked dots represent individual cases and the circles (triangles) indicate the mean value $\Delta_{\beta,\gamma}=0.5$ ($\Delta_{\beta,\gamma}=0.4$). The dotted line represents the probability of finding the optimal solution by random guessing and the dashed line represents the probability of finding the optimal solution if the search space is reduced by a quadratic speedup algorithm. (b) Approximation ratio for the W-MaxCut problems. The dashed line (GW) represents the approximation ratio using the best-known polynomial-time classical algorithm for the same problem. (c) TTS for SA and LR-QAOA compared to CPLEX's number of iterations. Three guiding lines are added following the perceived scale of the different algorithms. (d) 20 hard cases of Max-2-SAT problems (triangles) for problem sizes up to 20 qubits and 10 random cases of Max-3-SAT problems (circles) for problems up to 25 qubits. The inset shows the degeneracies of the first four eigenvalues for the Max-2-SAT problem highlighted in red. (e) 10 random MaxCut instances for problem sizes up to 35 qubits. The different colors mean the average percentage of edges compared to a fully connected graph. Therefore, darker blue, $0.05$, means a $5\%$ chance of having edges. The circles represent the mean value, and the dots represent the worst case. The inset shows the degeneracies of the solutions in the worst case. (f) The number of iterations needed to find the optimal solution using CPLEX for the MaxCut problems. The colors follow the same pattern as in (e).}
\end{figure*}

 In a previous version of this paper, we wanted to add further evidence that $probability(x^*) = 2^{-\eta N_q/p}$. Even though we consider that this is not the true scaling. The experiments are still useful to see how LR-QAOA behaves as the number of layers increases linearly with the problem size. We simulate LR-QAOA for fully connected W-MaxCut random problems for 100 cases and problem sizes ranging from 5 to 42 qubits using $p = N_q$. In Fig.~\ref{Fig:conjecture}, we show these cases. 
 
 Figure~\ref{Fig:conjecture}-(a) shows the $probability(x^*)$ for different problem sizes. The dots represent the probability of success for individual cases, and the black circle is the mean value for a $\Delta_{\gamma,\beta}=0.5$. 
After reaching $N_q = 25$, there is a decline in performance using $\Delta_{\gamma,\beta} =0.5$. To address this, we adjust $\Delta_{\gamma,\beta}$ to 0.4 for certain problem sizes, which leads to an improvement in performance. The data points corresponding to $\Delta_{\gamma,\beta}=0.4$ are shown with triangles. This suggests that one must reduce $\Delta_{\gamma,\beta}$ as the number of qubits increases. Interestingly, maintaining $N_q=p$ does not result in any significant performance loss.

Figure~\ref{Fig:conjecture}-(b) shows the approximation ratio for the same problems as (a). The decrement in performance is more notable in terms of the approximation ratio for $\Delta_{\gamma,\beta}=0.5$. The modification of $\Delta_{\gamma,\beta}=0.4$ leads to a general improvement in terms of the quality of the solutions. It has been proven by Theorem 4.2 in \cite{Hastad1997} that it is NP-Hard to approximate MaxCut for a factor of $16/17 \approx 0.941$  in the worst case. This indicates that the problem is hard to solve in the worst case, but not necessarily for the random instances presented here. The dashed line represents the minimum $r=0.878$ guarantee by the Goemans and Williamson (GW) algorithm~\cite{Williamson1994}, which is the best-known polynomial-time algorithm and, according to the unique games conjecture, also the best possible polynomial-time algorithm for this problem~\cite{Subhash2002, Khot2007}.

Figure~\ref{Fig:conjecture}-(c) shows the number of calls of the objective function, $TTS_{99\%}$, to reach a probability of finding the optimal solution with a 99\% probability, i.e., $p_d=0.99$ of the W-MaxCut problems for SA and LR-QAOA. We include the CPLEX number of iterations needed to find the optimal solution for comparison. This plot shows the average TTS for the 100 W-MaxCut cases.  We add three guiding lines of the perceived scaling of the 3 algorithms. CPLEX B\&B seems to scale $\approx 2^{0.36N_q}$, SA $\approx 2^{0.19N_q}$, and LR-QAOA $\approx 2^{0.11N_q}$. Note that we do not tune the $\Delta_{\gamma, \beta}$ parameters in these cases, and even though the results are promising, further improvement can be achieved by finding a function $\Delta_{\gamma, \beta}(N_q)$. In the case of LR-QAOA, at $N_q = 42$, it has improved by two orders of magnitude the TTS compared to SA and B\&B.

Fig.~\ref{Fig:conjecture}-(d) shows some instances of Max-2-SAT (triangles) and Max-3-SAT (circles) problems. The Max-3-SAT 10 random instances for problem sizes up to 25 qubits are chosen from a ratio of clauses/variables = 4.16, which is close to the critical region where the problems are known to be hard to solve \cite{Cheeseman1991, Zielinski2023}. In this case, the relation $probability(x^*)= 2^{-\eta N_q/p}$ seems to be valid on the average case. The Max-2-SAT problems used are obtained from a dataset of hard instances of this problem for SA and QA \cite{Mehta2021}. What makes these problems special is the number of degeneracies of the first excited state. For example, in the inset of this figure, it is shown a 20-qubit case is shown. The optimal solution has 2 degeneracies, while the first excited state has 6516. This causes SA and QA to get stuck in solutions close to optimal ones. In our case, LR-QAOA with $p=N_q$ does not hold a constant probability of success, so we can expect that $probability(x^*)= 2^{-\eta N_q/p}$ relation does not hold always. 

Fig.~\ref{Fig:conjecture}-(e) shows the LR-QAOA $p=N_q$ probability of success for 10 random cases of the MaxCut for up to 35 qubits. We use 7 different problems of MaxCut that we classify depending on the percentage of edges of the graph. The colors represent the average number of edges in the graph, meaning 0.05 having $5\%$ of all possible edges (dark blue) and $0.95$ a $95\%$ of all possible edges (dark red). In this case, on average the $p=N_q$ seems to hold $probability(x^*)= 2^{-\eta N_q/p}$. However, the worst case highlighted with the red circle deviates considerably from the average performance. We explore this case and find the same characteristic that makes Max-2-SAT hard for LR-QAOA, a high degeneracy close to the optimal solution. Finally, Fig.~\ref{Fig:conjecture}-(f) shows the number of iterations needed to solve the same MaxCut problems of Fig.~\ref{Fig:conjecture}-(e) using CPLEX. In this case, it is classically hard to find the optimal solution for densely connected graphs with a worse-than-quadratic speedup for edge probabilities from $80\%$ to $95\%$.

In the case of problems with suboptimal solutions densely concentrated close to the optimal, there is a noticeable exponential decay for $p=N_q$. In this scenario, for instance, in the Max-2-SAT problem, LR-QAOA still gives a quadratic speedup over random guessing. In this case, the high degeneracy causes the dense concentration of solutions close to the optimal one. Similarly, in the TSP or BPP, the concentration of solutions occurs due to penalization terms concentrating the valid solutions space on a small region. We suspect that in the limit, i.e., a random oracle with only one ground state and the first excited state containing all other possible solutions, like the one in \cite{Bennett1997}, one would need $p=2^{\eta N_q}$ layers to amplify the solution to some threshold for a constant $\eta$, similar to the performance guarantee in Grover's algorithm.

\subsection{Depolarizing Noise Aria QPU}\label{A:noise-aria}
\begin{figure*}
\centering
\includegraphics[width=17cm]{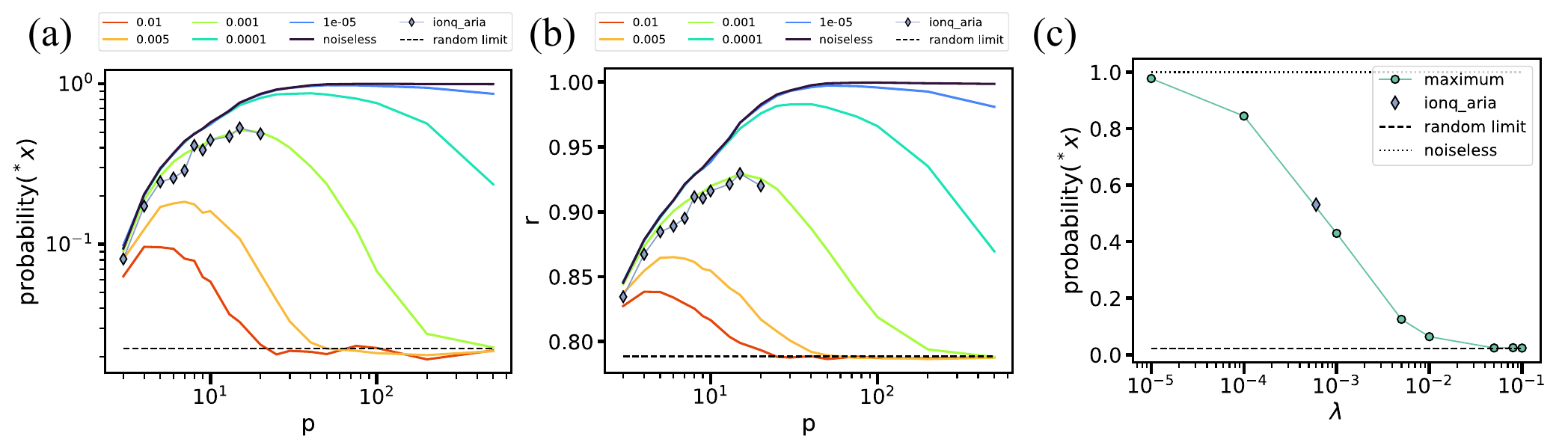}
\caption{\label{Fig:noiseionq} Results of LR-QAOA for the 10-qubit random WMaxcut problem at different $p$ when noise is added at different depolarizing noise strength $\lambda$. (a) probability of success (b) approximation ratio (c) largest probability of success for each $\lambda$. Diamonds represent the same problem executed on {\it ionq\_aria}. Different lines represent the strength $\lambda$ of the depolarizing channel. The shaded line shows the probability of obtaining the optimal solution by random guessing.}
\end{figure*}

Figure~\ref{Fig:noiseionq} shows 10-qubit WMaxcut problem results using LR-QAOA and depolarizing noise for $\lambda$s of Eq.~\ref{Eq:depolarizing} from $10^{-1}$ to $10^{-5}$ affecting only the 2-qubit gates. We compare these results with the ones obtained using {\it ionq\_aria} (diamonds) for the same problem. Fig.~\ref{Fig:noiseionq}-(a) shows how the probability of success vs.~the number of LR-QAOA layers for the different depolarizing noise strengths. Fig.~\ref{Fig:noiseionq}-(b) shows the approximation ratio vs.~the number of LR-QAOA layers for the same problem. In both plots, {\it ionq\_aria} results agree with a $\lambda = 6\times10^{-4}$. Finally, Fig.~\ref{Fig:noiseionq}-(c) shows the maximum probability of success vs.~$\lambda$. At $\lambda > 5 \times 10^{-2}$ information can still be recovered and at $\lambda = 10^{-5}$ errors are so slowly added that the algorithm can in some sense correct them, similar to what we show in Fig.~\ref{Fig:eig_evol}-(b) of the main text.

\subsection{Generalization of \( R_x(\theta) \) to Multiple Qubits} 
\label{A:UB_explained}
The single-qubit rotation around the x-axis is defined as
\begin{equation}
    R_x(\theta) = \exp\left(-j \frac{\theta}{2} X \right) = \cos\left( \frac{\theta}{2} \right) I - j \sin\left( \frac{\theta}{2} \right) \sigma^x.
\end{equation}
The matrix representation becomes
\begin{equation}
R_x(\theta) = 
\begin{bmatrix}
\cos\left( \frac{\theta}{2} \right) & -j \sin\left( \frac{\theta}{2} \right) \\
-j \sin\left( \frac{\theta}{2} \right) & \cos\left( \frac{\theta}{2} \right)
\end{bmatrix},
\end{equation}
to apply \( R_x(\theta) \) to both qubits in a 2-qubit system,
\begin{equation}
  R_x(\theta)^{\otimes 2} = R_x(\theta) \otimes R_x(\theta),  
\end{equation}

\begin{equation}
R_x(\theta)^{\otimes 2} =
\begin{bmatrix}
c^2 & -j c s & -j c s & -s^2 \\
-j c s & c^2 & -s^2 & -j c s \\
-j c s & -s^2 & c^2 & -j c s \\
-s^2 & -j c s & -j c s & c^2
\end{bmatrix}
\end{equation}

where $c = \cos(\theta/2)$ and $s = \sin(\theta/2)$. In general, for \( n \) qubits:
\[
R_x(\theta)^{\otimes n} = \underbrace{R_x(\theta) \otimes R_x(\theta) \otimes \cdots \otimes R_x(\theta)}_{n \text{ times}}
\]

This results in a \( 2^n \times 2^n \) unitary matrix. Let \( k, l \in \{0, 1, \dots, 2^n - 1\} \), and write their binary representations as $ k = (k_0 k_2 \dots k_{n-1})_2, \quad l = (l_0 l_2 \dots l_{n-1})_2
$. Then, the matrix entry at row \( k \), column \( l \) is
\begin{equation}\label{Eq:Rxn}
    \left[ R_x(\theta)^{\otimes n} \right]_{k,l} = \prod_{m=0}^{n-1} \left[ R_x(\theta) \right]_{l_m, k_m}.
\end{equation}  
There are two cases in the multiplication of Eq.~\ref{Eq:Rxn} when $k_m = l_m$, $\left[ R_x(\theta) \right]_{k_m,l_m} = c$, and $k_m \neq l_m$, $\left[ R_x(\theta) \right]_{k_m,l_m} = js$. Using the relation $k\cdot l = \sum_{m=0}^{n-1} k_m \oplus l_m$, Eq.~\ref{Eq:Rxn} can be rewritten as 

\begin{equation}
    \left[ R_x(\theta)^{\otimes n} \right]_{k,l} = c^{n-k\cdot l}(js)^{k\cdot l}.
\end{equation}
Therefore, in the evolution given by Eq.~\ref{UB} of the main text, $|\psi_{t+1}\rangle = U_B(\beta_t)|\psi_t\rangle$, each basis takes the form 

\begin{equation}
\alpha_k^{t+1} |k\rangle = \sum_{l=0}^{2^{N_q} - 1}\left[ R_x(2\beta_t)^{\otimes N_q} \right]_{k,l} \alpha_l^t |k\rangle= \sum_{l=0}^{2^{N_q} - 1}\cos(\beta_t)^{N_q-k\cdot l} (j\sin(\beta_t))^{k\cdot l} \alpha_l^t |k\rangle.
\end{equation}

\subsection{Max-3-SAT problem}
\label{Max-3-SAT}

The 3-SAT problem belongs to the family of Boolean satisfiability problems (SAT). 3-SAT has a specific structure called conjunctive normal form (CNF), which is described by 

\begin{equation}
\varphi = \bigwedge_{i}\big( \bigvee_j l_{i{_j}}\big)
\end{equation}

where $l_{i_{j}}$ are literals, i.e., any elements for the set of variables $x_k ~\forall ~ k \in \{0, ..., N_V-1\})$ or its negation $\neg x_k$, for $N_V$ variables. The $\vee$ and $\wedge$ represent the Boolean operations OR and AND, respectively. The terms $\big( \bigvee_j l_{i{_j}}\big)$ are called clauses, and in the case of 3-SAT, there are 3 literals for each clause. The total number of clauses is $N_C$.

Constructing the QUBO formulation for a individual clause, $C_i = (x_0\vee x_1 \vee x_2)$ is given by

\begin{equation}
(x_0\vee x_1 \vee x_2) = -(x_0 + x_1 + x_2 - x_0x_1 - x_0x_2 - x_1x_2+x_0x_1x_2),
\end{equation}
and the Hamiltonian representation of this QUBO in terms of the spin variables $x_i =  (1 + s_i)/2$ is given by

\begin{equation}\label{3SATQUBO}
(x_0\vee x_1 \vee x_2) = -\frac{1}{8}\left(s_0 + s_1 + s_2 - s_0s_1 - s_0s_2 - s_1s_2+s_0s_1s_2 + 7\right),
\end{equation}
The three-body interactions, $s_0s_1s_2$, in Eq.~\ref{3SATQUBO}, can be represented by the circuit shown in Fig. 10 of \cite{Jattana2023}. It allows the use of $N_q = N_V$ qubits to describe the problem, contrary to other QUBO formulations where it is needed $N_q = N_V + N_C$ variables \cite{Zielinski2023}. The Max-3-SAT version of the problem consists of finding the solution that maximizes the number of clauses satisfied.   

\subsection{Performance diagram}\label{sec:performance-diagram}
\begin{figure}[!tbh]
\centering
\includegraphics[width=8.5cm]{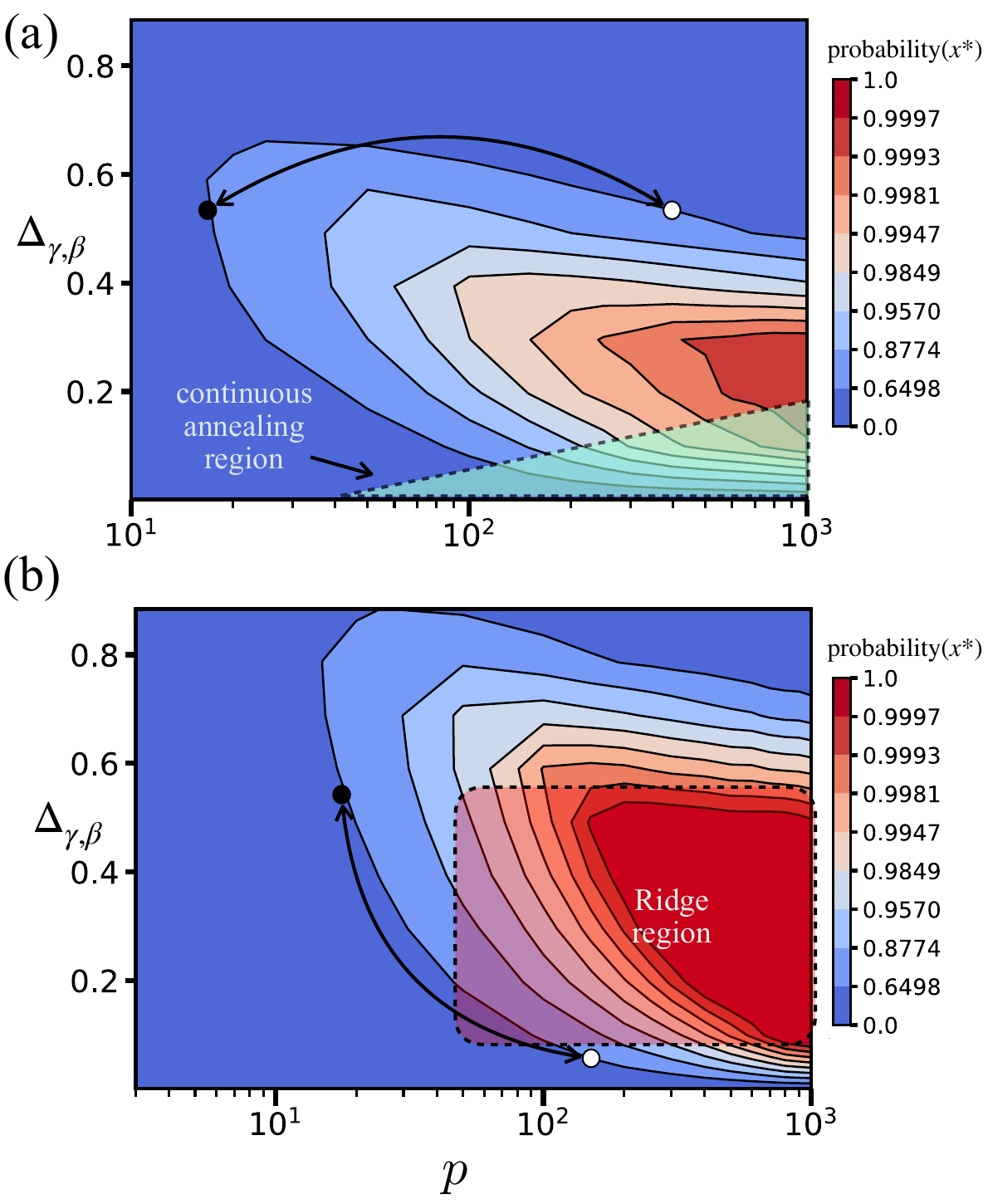}
\caption{\label{Fig:performance-diagram}Performance diagram using LR-QAOA on (a) a 10-qubit MIS and (b) a 10-qubit WMaxcut problem. The y-axis represents $\Delta_{\gamma,\beta}$, and the x-axis is the number of LR-QAOA layers. The colors represent the $probability(x^*)$ landscape, i.e., the probability of finding the optimal solution with darker red approaching the $100\%$. Circles in white and black represent similar $probability(x^*)$ cases. The arrows connecting them are guiding elements.}
\end{figure}

The performance diagram, introduced in \cite{Kremenetski2021}, is a clear way to visualize the LR-QAOA performance. Figure~\ref{Fig:performance-diagram}-(a) shows the performance diagrams of a 10-qubit MIS problem. In the green-shaded 'annealing' region, LR-QAOA behaves like a continuous QA protocol. This region is delimited by small $\Delta$ and large $p$, which guarantees yielding the optimal solution. The two points (black and white) marked in Fig. \ref{Fig:performance-diagram}-(a) show two LR-QAOA protocols with similar $probability(x^*)$. The white point is an LR-QAOA with $p=300$ while the black point shows an LR-QAOA of $p=17$.

In Fig.~\ref{Fig:performance-diagram}-(b), we present the performance diagram of a 10-qubit WMaxcut using LR-QAOA. The shaded red region in this plot illustrates the 'ridge' as defined by \cite{Kremenetski2021}. The ridge region is characterized by consistently improving performance as the number of layers grows. The two points in this plot show similar performance, with the black point achieving a $probability(x^*) > 0.65$ with just $p=20$, while the white point in the annealing region requires a higher $p=300$. In appendix \ref{Sec:2D-PD}, it is shown a 2-D version of the performance diagram for different problems.

\clearpage

\end{document}